\documentclass[12pt,preprint]{aastex}

\begin{document}

\newcommand{\blankline}{\par\vskip 12pt\noindent}
                     
\newcommand{\etastar}{\hbox{$\eta_A$}}
\newcommand{\etacar}{\hbox{$\eta$}~Car}

\newcommand{\EBMV}{\hbox{E$_{\hbox{\sc B-V}}$}}
          
\newcommand{\Mdot}{\hbox{$\dot M$}}
\newcommand{\Rsun}{\hbox{$R_\odot$}}
\newcommand{\Rstar}{\hbox{$R_*$}}
\newcommand{\Lsun}{\hbox{$L_\odot$}}
\newcommand{\Msun}{\hbox{$M_\odot$}}
\newcommand{\Msunyr}{\hbox{$M_\odot\,$yr$^{-1}$}}
\newcommand{\Teff}{\hbox{$T_{eff}$}}
\newcommand{\Vphot}{\hbox{$V_{\hbox{phot}}$}}
\newcommand{\Vcore}{\hbox{$V_{\hbox{core}}$}}
\newcommand{\Vinf}{\hbox{$V_\infty$}}
\newcommand{\Ne}{\hbox{$N_{\hbox{e}}$}}
\newcommand{\geff}{\hbox{$g_{\hbox{eff}}$}}
\newcommand{\sect}{\S}

\newcommand{\kms}{\hbox{km$\,$s$^{-1}$}}
\newcommand{\Hz}{\hbox{Hz}}
\newcommand{\mum}{\hbox{$\mu$m}}
\newcommand{\yr}{\hbox{yr$^{-1}$}}

\newcommand{\ie}{\hbox{i.e.,}} 
\newcommand{\eg}{\hbox{e.g.,}} 
\newcommand{\etal}{\hbox{et~al.}}
\newcommand{\etc}{\hbox{etc.}}
\newcommand{\Ang}{\hbox{\AA}} 

\newcommand{\Chil}{\hbox{$\chi^{\hbox{\sc L}}$}}
\newcommand{\Etal}{\hbox{$\eta^{\hbox{\sc L}}$}}

\newcommand{\etavec}{\hbox{$\bf \eta$}}
\newcommand{\chivec}{\hbox{$\bf \chi$}}
\newcommand{\Jvec}{\hbox{\bf J}}
\newcommand{\Hvec}{\hbox{\bf H}}

\newcommand{\LX}{\hbox{L$_{\hbox{X}}$}}

\newcommand{\NF}{\hbox{$^{F}n$}}
\newcommand{\NS}{\hbox{$^{S}n$}}
\newcommand{\Nlev}{\hbox{$N_{\hbox{lev}}$}}

\title{The UV Scattering Halo of the Central Source 
Associated with Eta Carinae}

\author{D. John Hillier}
\affil{Department of Physics and Astronomy,
University of Pittsburgh, Pittsburgh, PA 15260}
\and
\author{T. Gull}
\affil{Exploration of the Universe, 
NASA/Goddard Space Flight Center, Code 667, Greenbelt, MD 20771}
\and
\author{K. Nielsen}
\affil{Department of Physics, Catholic University of
American Washington DC \& Exploration of the Universe,  
NASA/Goddard Space Flight Center, Code 667, Greenbelt, MD 20771}
\and
\author{G. Sonneborn}
\affil{Laboratory for Observational Cosmology,
NASA/Goddard Space Flight Center, Code 665, Greenbelt, MD 20771}
\and
\author{R. Iping}
\affil{Department of Physics, Catholic University of
American, Washington DC 20064 \&
NASA/Goddard Space Flight Center, Code 665, Greenbelt, MD 20771}


\and
\author{Nathan Smith}
\affil{Center for Astrophysics and Space Astronomy, University of Colorado, 389 UCB,
Boulder, CO 80309}

 
\and
\author{M. Corcoran}
\affil{Laboratory of High Energy Astrophysics,
NASA/Goddard Space Flight Center, Code 662, Greenbelt, MD 20771}

\and
\author{A. Damineli}
\affil{Instituto de Astronomia, Geofisica e 
Ciencias Atmosfericas da USP, R. do Mat\~ao, 05508-900, S\~ao Paulo, Brazil}

\and
\author{F. W. Hamann}
\affil{University of Florida}

\and
\author{J. C. Martin}
\affil{Department of Physics and Astronomy,
University of Minnesota}

\and
\author{K. Weis\footnote{Lise-Meitner fellow}}
\affil{Astronomisches Institut, Ruhr-Universitaet Bochum, 
Universitaetsstr. 150 44780 Bochum, Germany}

%
 

\altaffiltext{1}{Based on observations with the NASA/ESA Hubble Space
     Telescope, obtained at the Space Telescope Science Institute,
     which is operated by the Association of Universities for Research
     in Astronomy, Inc., under NASA contract NAS5-2655.}

\begin{abstract}
We have made an extensive study of the UV spectrum of $\eta$ Carinae, and
find that we do not directly observe the star and its wind in the UV. Because of
dust along our line of sight, the UV light that we observe
arises from bound-bound scattering at large impact parameters (e.g.,
0.033\arcsec). We obtain a reasonable fit to the
UV spectrum by using only the flux that originates outside
0.033\arcsec. This explains why we can still observe
\etastar\ in the UV despite the large optical extinction ---
it is due to the presence of an intrinsic coronagraph in
the $\eta$ Carinae system, and to the extension of the UV 
emitting region. It is not due to peculiar dust properties alone. 
We have computed the spectrum of the purported companion
star, and show that it could only be directly detected
in the UV spectrum preferentially in the 
Far Ultraviolet Spectroscopic Explorer (FUSE) spectral
region (912-1175\,\AA). However, we find no direct evidence for
a companion star, with the properties indicated by X-ray studies
and studies of the Weigelt blobs,
in UV spectra. This might be due to
reprocessing of the companion's light by the dense stellar
wind of the primary. FUSE observations, at epochs
when (and if) the opening of the bow shock is along
our sightline, should have a better chance of detecting the
companion spectrum.
The UV spectrum is dominated
by low ionization lines, many exhibiting P~Cygni profiles. 
Some of the strongest lines that 
can be readily identified include \ion{C}{2} $\lambda$ 1335 (UV1),
\ion{Si}{2}  $\lambda\lambda$1304, 1309 (UV3); $\lambda$ 1264 (UV4), 
$\lambda\lambda$ 1527, 1533 (UV 2); $\lambda\lambda$  1808, 1817 (UV1),
\ion{S}{2} $\lambda\lambda$ 1250, 1253 (UV1), 
\ion{Al}{2} $\lambda$1671, 
\ion{N}{1}\ $\lambda\lambda$ 1493, 1495 (UV4), 
\ion{Mg}{2} $\lambda\lambda$ 2796, 2803, as well as numerous \ion{Fe}{2} lines.
Higher excitation lines due to \ion{Al}{3} $\lambda\lambda$1855, 1863
and \ion{Si}{4} $\lambda\lambda$1394, 1403 
can be identified.  A previous identification of \ion{C}{4}
$\lambda\lambda 1548, 1552$ must, because of severe blending,
be considered as uncertain. The terminal velocity, as derived from
numerous emission lines, is less than 600\,\kms, with preferred values
around 520\,\kms. This value is consistent with
that seen in optical spectra.
Broad \ion{Fe}{2} and [\ion{Fe}{2}] emission lines are detected in
spectra taken in the SE lobe, 0.2\arcsec\ from the central star.
These lines arise in the stellar wind --- thus the Space Telescope
Imaging Spectrograph (STIS) and the Hubble Space Telescope (HST)
are resolving, at some wavelengths, the stellar wind of
Eta Carinae.  The wind spectrum shows some similarities to the spectra of
the B \& D Weigelt blobs, but also shows some marked differences
in that high excitation lines, and lines pumped
by Ly$\alpha$, are not seen.
The resolution of the stellar wind at optical wavelengths,
and the detection of the broad lines, lends support to 
our interpretation of the UV spectrum,  and to our model for $\eta$ Carinae. 

\end{abstract}


\section{Introduction}
 
Eta Carinae is one of the most luminous and spectacular stars in our galaxy,
and exhibits variability on a wide range of time scales
(Zanella, Wolf, \& Stahl \citeyear{ZWS84_shell},
Davidson and Humphreys \cite{DH97_rev}). A major breakthrough to
understanding some of the variability was made by Damineli \cite{Dam96_cyc}
who found that nebular line strengths varied periodically on 
a time-scale of 5.54 years. A similar time-scale was also present in infrared 
data (Whitelock \etal\ \citeyear{WFK94_eta_var}). The cycle was confirmed to
be periodic when the event predicted for December 1997 came on schedule
(Damineli \etal\ \citeyear{DKW00_bin}, Feast \etal\ \citeyear{FWM01_IRvar}). 
Variability seen in radio data can also be related to a similar time-scale
(Duncan \etal\ \citeyear{DWRL99_eta_radio}, Duncan \& White \citeyear{DW03_eta_radio}). 

On the basis of radial velocity variations, Damineli, Conti \& Lopes (\citeyear{DCL97_eta_bin})
postulated that Eta Carinae is a binary system characterized by high eccentricity, 
a hotter companion, and strong colliding winds.
X-ray observations, which reveal an apparent X-ray eclipse, appear to
confirm the binary hypothesis (Ishibashi \etal\ \citeyear{ICD99_xray}, 
Corcoran \etal\ \citeyear{CIS01_xray}). The duration ($\sim3$ months)
of the eclipse, together
with the rapid variability of high excitation nebular lines, indicates
that the orbit is highly elliptical, with an eccentricity greater than 0.8
(see, e.g., Corcoran \citeyear{CIS01_xray}).  However, fitting the long 
duration of the X-ray minimum has been a challenge that was tentatively 
circumvented by enhanced mass-loss at periastron (Corcoran et al. \citeyear{CIS01_xray}) 
or with a tilted angle of the colliding wind shock cone 
(Pittard \& Corcoran \citeyear{PC02_xray}, Ishibashi \citeyear{Ish01_orbit}).
In addition, radiative transfer effects and 
the complicated and severely blended profiles observed from the ground
make it difficult to measure, and to interpret, radial velocity measurements.
These difficulties led Davidson \cite{Dav99_bin} to question the 
validity of the derived orbital parameters. Indeed, high spatial resolution 
observations with the Space Telescope
Imaging Spectrograph (STIS) did not reveal the expected velocity shifts in emission 
lines associated with the primary star (Davidson \etal\ \citeyear{DIG00_bin}).
An alternative set of orbital parameters, based on analysis of the
X-ray light curve, has been given by Ishibashi \cite{Ish01_orbit}.
Thus, while the binary model is generally accepted, the nature of the companion star, 
its orbit, and the influence of the companion on the major outbursts of the 1840s and 
1890s are uncertain. Other interpretations of the variability have been suggested,
most notably shell ejections (e.g., Davidson \etal\ \citeyear{DMH05_Ha},
Martin \etal\ \citeyear{MDH05_4686}). A combination of shell ejections and
binarity might be needed to explain variability seen in 
{\it Hubble Space Telescope (HST)} data
(e.g., Smith \etal\ \citeyear{SDG03_eta_lat}). 

Information on the nature of the companion comes primarily
from indirect arguments. Analysis of X-ray data indicates that the 
companion should have a mass loss rate of approximately 
$\Mdot=1.0 \times 10^{-5}$\,\Msunyr, while estimates of the terminal
velocity range from 1700\,\kms\ (Corcoran \etal\ \citeyear{CIS01_xray}) to
3000\,\kms\ (Pittard \& Corcoran \citeyear{PC03_xray}), with
the higher velocity estimates now preferred. Other constraints
on the companion come from its influence on the spectrum of
the Weigelt blobs (Verner \etal\ \citeyear{VGB02_bd_blob}; Verner,
Bruhweiler \& Gull \citeyear{VBG05_bd_blob}). These suggest that it is
an O type star with an effective temperature between 34,000\,K and 38,000\,K.

With the aim of clarifying the nature of the 5.5 year periodicity, and 
to gain insights into both the nature of the primary and companion stars,
we initiated a major observational multi-wavelength campaign, 
using the Hubble Space Telescope
(HST Eta Carinae Treasury project; PI K. Davidson),
FUSE (PI T. Gull), X-ray satellite observatories (PI. M. Corcoran), 
and numerous ground-based observatories (e.g.,
VLT-UVES; PI. K. Weis), 
to observe Eta Carinae through a full variability cycle.
Some of the HST data used in this
paper are based on data obtained as part of that campaign, while other
data were obtained as part of earlier HST programs to understand Eta Carinae.
The campaign has confirmed the 5.5 year periodicity
(Whitelock \etal\ \citeyear{WFM04_event}, Corcoran \citeyear{Cor05_xray_mon}, 
White \etal\ \citeyear{WDC05_radio}) and provided a wealth of new 
data to help understand the complex system that is Eta Carinae.

An introduction to the extensive literature on Eta Carinae can be obtained
from the reviews by Humphreys \& Davidson \cite{HD94_LBV_rev}, and
Davidson \& Humphreys \cite{DH97_rev}, and three
relatively recent workshops devoted to Eta Carinae and related objects
(Morse, Humphreys, \& Damineli \citeyear{Eta_wrk_99};
Gull, Johansson, \& Davidson \citeyear{Hven_wrk_shop};
Humphreys \& Stanek \citeyear{GT05}).

\subsection{The Primary Star: \etastar}

Ground-based spectra of Eta Carinae reveal a complex spectrum 
of H, \ion{He}{1}, \ion{Fe}{2}, and [\ion{Fe}{2}] 
emission lines with 2 principal
components. There is a narrow nebular-like spectrum $(V_{\hbox{FWHM}}
< 40\,\kms)$, and a broad spectrum which indicates gaseous outflows
with a velocity of approximately 500\,\kms. As other components are also
seen, it is difficult to discern the underlying nature of the primary star, \etastar.
Using the reflected spectrum Hillier \& Allen \cite{HA92_eta} suggested that the
spectrum of \etastar\ is similar to the extreme P~Cygni star
HDE\,316285. Later observations with the HST confirm this
suggestion (Davidson \etal\ \citeyear{DEW95_FOS}; 
Hillier \etal\ \citeyear{HDIG01} [hereafter HDIG]).
With the STIS on the HST it is possible to obtain the spectrum of the 
central star, uncontaminated by lines from the adjacent 
Weigelt blobs and the Homunculus. 
The Weigelt blobs are spatially unresolved condensations first seen in
ground based speckle studies of $\eta$ Carinae
(Weigelt, \& Ebersberger \citeyear{WE86_blobs}; 
Hofmann \& Weigelt \citeyear{HW88_blobs}).
It is the Weigelt blobs that give rise to the narrow nebular
spectrum (Davidson \etal\ \citeyear{DEW95_FOS},
\citeyear{DEJ97_UV_blobs}). The Little Homunculus also contributes to the
narrow nebular spectrum (Ishibashi \etal\ \citeyear{Ish03_lit_hom};
Smith \citeyear{Smi05_lit-hom}).

HDIG were able to model the March 1998 optical spectrum, taken
a few months after the 1997 spectroscopic event 
(the X-ray minimum began in mid-December of 1997;
see, e.g., Corcoran \citeyear{Cor05_xray_mon}) using CMFGEN --
a non-LTE line blanketed code designed for modeling stars
with stellar winds (Hillier \& Miller 
\citeyear{HM98_blank}, \citeyear{HM99_WC}).
The code assumes spherical symmetry, and that the star and
its wind is spatially unresolved.
They were able to obtain a good fit to the H, \ion{He}{1},
and \ion{Fe}{2}\ emission line spectrum using a 
luminosity of $5 \times 10^6$\,\Lsun\ for the primary star,
and assuming a mass-loss rate of $1.0\times 10^{-3}$\,\Msunyr\ 
and solar mass-fraction of iron. The best fit model confirmed that the
central star suffered, in March 1998, severe circumstellar extinction with a total
visual extinction of 7 magnitudes. The primary difference between
the March 1998 spectrum, and later spectra, is that the P~Cygni
profiles tend to be stronger, and more prevalent in the March 1998 data
set. In the optical,
most emission line strengths (i.e, EWs) are 
very similar to those obtained on 4-Jul-2002 --- indeed the
two data sets are in better agreement with each other than
with the CMFGEN model.
The major exceptions are the \ion{He}{1}\ profiles which show
significant profile variations. Surprisingly, there is no strong 
indication that the terminal velocity of the outflow has changed.
Changes in the H$\alpha$ profile with time have been discussed by Davidson
\etal\, \cite{DMH05_Ha}, while other changes will be the subject
of future papers. 

While the fit to the emission line spectrum was satisfactory 
(but see HDIG for more details), there were two fundamental
discrepancies between the best fit model and the observations. First, the
models predicted much stronger absorption components associated
with  \ion{H}{1}, \ion{He}{1}\ and \ion{Fe}{2} P~Cygni emission lines than was observed.
Second, the fit to the spectrum shortward of $1600\,$\AA\ was very poor
-- emission features did not usually match and
the UV spectrum was much less absorbed, by the wind, than predicted by the model.
Since the emission lines sample the whole wind, but the wind absorption
components sample only one line of sight, two possible causes were
suggested. First, the discrepancies could arise because Eta's wind is 
asymmetric. This would not be surprising since the Homunculus is bipolar,
and other LBVs, such as AG~Car, are known to possess asymmetric winds
(e.g., Schulte-Ladbeck \citeyear{SCH94_AG}).
For this to work, the wind would have to be more ionized
along our sightline. Recent observations of the reflected optical
spectrum by Smith \etal\ \cite{SDG03_eta_lat} provide direct evidence that the
wind is aspherical and probably axisymmetric (bipolar),
with stronger Balmer absorption near the poles.  Interferometric observations 
by van Boekel \etal\ \cite{Boek03}, confirm the bipolar geometry of the wind. 
Second, the ionization of the wind in some regions could be influenced by the
ionizing radiation field of the companion. This is appealing since
it provides a simple explanation for some of the observed spectral changes.

In this paper we reexamine the formation of the UV spectrum.
We identify major wind lines seen in UV spectra, 
discuss the terminal velocity
of the wind, and estimate the intervening H column density.
Extensive foreground absorptions, caused by both circumstellar
and interstellar matter, strongly influence the UV spectrum
and have been discussed elsewhere (Gull \etal\ \citeyear{GVB05_abs};
Nielsen, Gull, \& Vieira Kober, \citeyear{NGK05}). 
We investigate the reason for our model's failure to
explain the observed UV spectrum. Several alternatives are
considered including wind asymmetries, ionization of the
outer wind by a companion star, the direct influence
of a companion spectrum, the influence of dust, and the
spatial extension of the UV emitting region. We show that the 
original model, when extended and interpreted differently,
can explain many of the features seen in the UV spectrum.

The paper is organized as follows: In
\S \ref{Sec_obs} we discuss the observations and data reduction
while in \S \ref{Sec_dist_red} we examine the distance, visual magnitude 
and reddening of \etastar, the primary star associated with
$\eta$ Carinae. Sections \ref{Sec_nat_UV},
\ref{Sec_line_id} and \ref{Sec_vinf} discuss
the nature of the UV spectrum, the identification of UV wind lines,
and the terminal velocity of the wind.
The spectrum of the companion star, and its possible influence
of the observed spectrum is discussed in \S \ref{Sec_companion_star}.
In \S \ref{Sec_fuse} we examine the FUSE spectrum, while the
\ion{H}{1}\ column density along our sight line is derived in
\S \ref{Sec_HI_col}.
The Model used for the analysis, and improvements made to it for the
UV analyses, are described in \S \ref{Sec_model}, while the possible
importance of wind asymmetries and flow times are discussed in
\S \ref{Sec_asym}. The creation of the UV spectrum is discussed in 
\S \ref{Sec_UV_mystery}. Finally, in \S \ref{Sec_out_wind_spec} we
discuss how STIS observations allow us to study the spatial structure
of the outer wind at optical wavelengths.

\section{Observations and Data Reduction} 
\label{Sec_obs}

Here we are concerned primarily with UV spectral observations.
UV images ($\lambda > 2100$\,\AA) were discussed recently by Smith \etal\ 
(\citeyear{SMC04_UV2}, \citeyear{SMG04_UV1})
while optical and IR spectra of the star have been discussed by
HDIG; Smith \cite{Smi02_IR_Image}; Hamann \etal\ \cite{HDJ94_eta_opt}; 
Hillier \& Allen \cite{HA92_eta};
Allen, Jones, \& Hyland, \cite{AJH85_IR}.
More recently, the variable H$\alpha$ profiles have been discussed by
Davidson \etal\ \cite{DMH05_Ha}, while the discovery and interpretation of
the \ion{He}{2}\ $\lambda$4686 line has been discussed by
Steiner \& Damineli \cite{SD04_4686}, Gull \cite{Gull05}, 
Stahl \etal\ \cite{SWB05_eta_event} and
 Martin \etal\ \cite{MDH05_4686}.
Using reflected spectra, the latitude dependence of the wind variations
in optical lines has been studied by Weis \etal\ \cite{WSB05_polar_var}
and Stahl \etal\ \cite{SWB05_eta_event}.

The HST/STIS UV observations were recorded with MAMA echelle modes
between March 2000 and March 2004. 
These data are part of an extensive optical and UV data set obtained
as part of several HST programs including
the HST treasury program on $\eta$ Carinae (PI: K. Davidson, GO-9420
and GO-9973).
A summary of the UV MAMA observations, including
PI and proposal number, is provided in Table~\ref{tab_uv_obs}.
The E140M and E230H modes were used to
span the spectral range from 1175 to 2380\,\AA\ with 30,000 to 60,000
resolving power. As instrument sensitivity and source brightness
combined to provide better detectivity, the E230H mode was used to span
2380 to 3160\,\AA\ at resolving power 100,000. 
Data reduction was accomplished using the STIS GTO IDL CALSTIS
software (Lindler 2002) with a special modification for large aperture
with nebular extended source to correct for nebular background.


A summary of the optical observations is provided in Table~\ref{tab_opt_obs}.
The data set is very extensive, although not all observations were done in the
same manner. Different slit orientations, imposed by spacecraft orientations,
were used, and on some dates only limited wavelength coverage was obtained. 
For most of the analyses (and plots) in this paper we have used the UV data 
obtained in July 2002 although data from other epochs has also been studied.
The extensive July 2002 data data set is during Eta Carinae's broad maximum in 
its 5.54 year period. 


\begin{table*}
\caption[]{Summary of UV Observations of Eta Carinae}
\label{tab_uv_obs}
\begin{center}
\begin{tabular}{llllll}
\hline
Date          &HST &  Principal & Aperture  &    Grating        &     Spectral Range      \\
              & ID & Investigator & ( \arcsec) & & (\AA) \\
\hline
2000 Mar 23 &   8327   &       K. Davidson    &         0.2x0.2  &   E140M, E230M    &          1175-2360       \\
2001 Oct 1 &  9242    &      A. Danks        &         0.2x0.09 &     E230H         &        2385-2943       \\
2002 Jan 20 &  9083    &      K. Davidson     &         0.2x0.09 &     E230H         &        2886-3159       \\
              &          &                      &         0.2x0.2  &   E140M, E230M    &      1175-2360       \\
2002 Jul 4 &  9337    &      K. Davidson     &        0.2x0.2   &    E140M, E230M   &             1175-2360       \\
              &          &                      &                  &   E230H           &        2423-2596       \\
2003 Feb 13 &   9420   &       K. Davidson    &         0.2x0.2  &   E140M, E230M    &           1175-2360       \\
2003 May 26 &    9420  &        K. Davidson   &         0.2x0.2  &  E140M, E230M     &         1175-2360       \\
2003 Jun 1 &  9420    &      K. Davidson     &         0.2x0.2  &    E140M, E230M   &             1175-2360       \\
2003 Jun 22 &  9420    &      K. Davidson     &         0.2x0.2  &    E140M, E230M   &             1175-2360       \\
2003 Jul 5 &  9973    &      K. Davidson     &         0.2X0.2  &   E140M, E230M    &           1175-2360       \\
2003 Jul 29 &   9973   &       K. Davidson    &        0.3x0.2   &  E140M, E230M     &          1175-2360       \\
              &          &                      &                  &  E230H            &          2385-3159       \\
2003 Sep 21 &   9973   &       K. Davidson    &        0.3x0.2   &  E140M, E230M     &          1175-2360       \\
              &          &                      &                  &   E230H           &           2385-3159       \\
2004 Mar 6 &  9973    &      K. Davidson     &        0.3x0.2   &   E140M, E230M    &            1175-2360       \\
              &          &                      &                  &    E230H          &           2385-3159       \\
\hline
\end{tabular}
\end{center}
\end{table*}

\begin{table*}
\caption[]{Summary of Optical Observations of Eta Carinae}
\label{tab_opt_obs}
\begin{center}
\begin{tabular}{llllll}
\hline
Date          &HST &  Principal & Aperture  &    Grating        &     Spectral Range      \\
              & ID & Investigator & (\arcsec) & & (\AA) \\
\hline
1997 Dec 31 &    7302   &       K. Davidson    &      52x0.1    & G230MB, G430M, G750M   &     Selected       \\
1998 Mar 19 &    7302   &        K. Davidson   &       52x0.1   & G230MB, G430M, G750M   &   1640-10300       \\
1998 Nov 25 &    8036   &       T. Gull        &       52x0.1   & G230MB, G430M, G750M   &    Selected       \\
1999 Feb 21 &    8036   &       T. Gull        &       52x0.1   & G230MB, G430M, G750M   &     1640-10100       \\
2000 Mar 13 &    8327   &       K. Davidson    &      52x0.1    & G750M                  &      6768       \\
2000 Mar 20 &    8483   &       T. Gull        &       52x0.1   & G230MB, G430M, G750M   &     1640-10100       \\
2000 Mar 21 &    8483   &       T. Gull        &       52x0.1   & G430M, G750M           &     4961, 6768 Mapping       \\
2000 Oct 9 &     8327   &      K. Davidson     &     52x0.1     & G750M                  &       6768       \\
2001 Apr 17 &    8619   &       K. Davidson    &      52x0.1    & G230MB, G430M, G750M   &      1640-10100       \\
2001 Oct 1 &   9083    &      K. Davidson     &     52x0.1     & G230MB, G430M, G750M   &       Selected       \\
2001 Nov 27 &    8619   &       K. Davidson    &      52x0.1    & G750M                  &      6768       \\
2002 Jan 19 &    9083   &       K. Davidson    &      52x0.1    & G230M, G430M, G50M     &      1640-10100       \\
2002 Jan 20 &    9083   &       K. Davidson    &      52x0.1    & G750M                  &      6768       \\
2002 Jul 4 &   9337    &      K. Davidson     &     52x0.1     & G230MB, G430M, G750M   &      1640-10100       \\
2002 Dec 16 &    9420   &       K. Davidson    &      52x0.1    & G750M                  &      6768       \\
2003 Feb 12 &    9420   &       K. Davidson    &      52x0.1    & G230MB, G430M, G750M   &      1640-10100       \\
2003 Feb 13 &    9420   &       K. Davidson    &      52x0.1    & G750M                  &        6768       \\
2003 Mar 29 &    9420   &       K. Davidson    &      52x0.1    & G230MB, G430M, G750M   &      Selected       \\
2003 May 5 &   9420    &      K. Davidson     &     52x0.1     & G230MB, G430M, G750M   &       Selected       \\
2003 May 17 &    9420   &       K. Davidson    &      52x0.1    & G230MB, G430M, G750M   &      1640-10100       \\
2003 May 26 &    9420   &       K. Davidson    &      52x0.1    & G750M                  &        6768       \\
2003 Jun 1 &   9420    &      K. Davidson     &     52x0.1     & G750M                  &         6768       \\
2003 Jun 22 &    9420   &       K. Davidson    &      52x0.1    & G230MB, G430M, G750M   &      1640-10100       \\
2003 Jul 5 &   9973    &      K. Davidson     &     52X0.1     & G230MB, G430M, G750M   &       1640-10100       \\
2003 Jul 29 &    9973   &       K. Davidson    &      52x0,1    & G750M                  &          6768       \\
2003 Jul 31 &    9973   &       K. Davidson    &      52x0.1    & G230MB, G430M, G750M   &      1640-10100       \\
2003 Sep 22 &    9973   &       K. Davidson    &      52x0.1    & G230MB, G430M, G750M   &      1640-10100       \\
2003 Nov 17 &    9973   &       K. Davidson    &      52x0.1    & G230MB, G430M, G750M   &      Selected       \\
2004 Mar 6 &   9973    &      K. Davidson     &     52x0.1     & G230MB, G430M, G750M   &       Selected, 2500-10100 \\
\hline
\end{tabular}
\end{center}
\end{table*}


Three sets of FUSE FUV spectra, which cover the wavelength band
990 to 1187\,\AA, were utilized for the modeling.
Two sets of FUSE observations were obtained with the LWRS
($30\arcsec\times30\arcsec$) aperture on 2002 June 25 (obsID C1540101), and
2003 June 10 (obsID D0070102). These and other FUSE observations of
$\eta$ Carinae are discussed in more detail by Iping et al. (2005
and in preparation). The second of these two exposures was taken
approximately 20 days before the onset of X-ray minimum (2003 June 29).
The exposure times were 29157 and 15282 seconds, respectively. The
standard CalFUSE calibration pipeline data products were used.
The individual exposures in each observation were aligned by
cross-correlation and coadded. The 2002 and 2003 spectra are
qualitatively similar,
but do show significant variations in the strength of absorption and emission
features. After the observations were obtained it was realized that
two 11$^{th}$ magnitude B-type stars, located 13\farcs9 from $\eta$
Carinae, could contaminate the LWRS spectra
by $\sim$50\% (Iping et al. 2005). The third FUSE spectrum, obtained
on 2004 April 11 (obs. ID: D0070109),
using the $1\farcs25\times20\arcsec$ LiF1 HIRS aperture at a PA of
$\sim$134$\arcdeg$ with an exposure time 17118 seconds,
shows significant differences with the LWRS observations.
The HIRS data were
processed with the same techniques as the LWRS spectra. An
additional correction, for the point-source throughput of the HIRS aperture (60\%),
was not made to the spectrum used in this paper.
This would raise the flux level of the HIRS spectrum by a factor of 1.67. 
As shown by Iping et al. (2005), the 2004 HIRS observation probably
represents the intrinsic FUV spectrum of $\eta$ Carinae and there is little FUV
flux that arises outside the HIRS aperture, except for the two B
stars. Because it is free of contamination, all comparisons between models
and the FUV spectrum in this paper are made with the HIRS spectrum.

\section{Distance, Visual Magnitude and Reddening}
\label{Sec_dist_red}

\subsection{Distance}
Eta Carinae is located in a region of massive star formation
in the Carina nebula, and is associated with the massive cluster
Trumpler 16 (Walborn 
\citeyear{Wal73}). The distance of the Carinae nebula is generally taken to
be around 2.5\,kpc, however for Eta Carinae it is possible to derive
an accurate estimate of the distance using the Homunculus.
Using the explosion and basic
geometrical arguments, Allen \& Hillier (\citeyear{AH93_hom}) derived
a distance of $2.2\pm 0.2$\,kpc, while Davidson and Humphreys \cite{DH97_rev}
using several different arguments, obtained 2.3\,kpc. 
More recently Davidson \etal\
(\citeyear{DSG01_hom_shape}) derived $d=2.25 \pm 0.18$\,kpc while 
Smith (\citeyear{Smi02_IR_Image}) found that a distance of 2.25\,kpc 
yielded images of the Homunculus with the greatest degree of
axial symmetry. Thus based on the best available estimates, and for 
consistency with earlier work, we have adopted $d=2.3$\,kpc. 

\subsection{Reddening and Visual Magnitude}
\label{Sec_vmag}

The reddening law towards Eta Carinae is known to be unusual.
Viotti \etal\ (\citeyear{VRC89_UV}), for example,  suggested that the central
source suffers an additional 0.7 magnitudes of color excess
over the color excess of E(B-V)$=0.4$ derived from the depth of the
interstellar 2200\AA\ band. In fact, given the heavily reddened
optical spectrum, it is surprising that the UV spectrum is easily detected.
This suggests that the circumstellar reddening law must be flat.
Detailed studies of the stellar spectrum (1700\AA\ to 10000\AA) confirm the
unusual reddening law (e.g., HDIG), and thermal-IR emission indicates dust temperatures
consistent with large grains, which could cause unusual reddening
(Smith \etal\ \citeyear{SGK98_IR_morph}, \citeyear{SGH03_IR}). 
Using detailed modeling, HDIG determined that the visual extinction in March 1998 was
7 magnitudes. The wavelength dependence was unusual and was fitted
using the Cardelli, Clayton  \& Mathis (\citeyear{CCM88_ext}) extinction law assuming
R=5.0 and E(B-V)=1. In addition, there were 2 magnitudes of
gray extinction. Most of the extinction must arise from the circumstellar
material associated with $\eta$ Carinae, and there is evidence that this
extinction is variable with time.

STIS observations reveal that \etastar\ has brightened by about
a factor of 3 between 1998.0 and 2003.7 
although the increase has not been uniform in time
(Davidson \etal\ \citeyear{DGH99_eta_bright}, Martin \etal\ \citeyear{MK04_eta_var}).
The best photometric data set on the brightening of the central source
was obtained using the STIS acquisition images which utilizes a neutral
density filter, and which covers the wavelength regions from
2000 to 11000\AA.\footnote{Comparison of the STIS spectroscopic
data of 1998-March-19 data with the data of 2002-July-04 (2003-June-22) shows that
the brightening was somewhat larger in the near UV than the optical
 --- roughly a factor of 1.5 (2) near 10000\AA, increasing to roughly a factor
of 2.5 (3) at 2500\AA.}
Multiepoch HST/WFPC2 images of eta Carinae also reveal a brightening of the homunculus,
although the behavior relative to the central star is complex
(Smith \etal\ \citeyear{SMDH00_Hbright}).
Analysis of the STIS observations shows that the increase
has also occurred in the UV. For example, between 2000-March-23 and
2003-June-22 the UV flux (averaged over the wavelength
interval 1250\,\AA\ to 1700\,\AA)
increased by roughly a factor of 1.7. With the exception of observations
around the event, both optical and UV spectra suggest that there has been
very little change in the excitation temperature inferred from the
wind emission spectrum. This implies that the effective temperature
of the underlying star has not changed. Unless the bolometric luminosity 
has increased, and the star has conspired to alter its mass loss
but not the wind terminal velocity to preserve its spectrum, this flux change
can only be interpreted as due to a decrease in circumstellar 
extinction.
This interpretation in also consistent with observations of the
Weigelt blob spectra. Direct comparison of the Weigelt Blob D spectrum from March 1998 to
September 2003 (5.5 years, or one period, apart) shows they are
essentially identical in both line strengths and fluxes.
In ground-based observations, the strength of the
nebular (blob) emission lines, relative to the continuum, has been declining with time 
(e.g., Damineli, Levenhagen, \& Leister \citeyear{DLL05}).
We estimate that the visual extinction has declined from
approximately 7 magnitudes in March 1998 to approximately 6.3 magnitudes
in July of 2002.  Both Davidson \etal\ \cite{DGH99_eta_bright} and
Martin \etal\ \cite{MK04_eta_var} provide a detailed discussion
and insight into the brightening of \etastar.

\section{The Nature of the UV Spectrum}
\label{Sec_nat_UV}


The IUE spectrum of Eta Carinae has been studied by many different groups
(e.g., Viotti \etal\ \citeyear{VRC89_UV}; Viotti \& Rossi \citeyear{VR99_IUE};
Ebbets, Walborn \& Parker \citeyear{EWP97_eta_UV}).
From these, and other studies, it is known that
the UV spectrum is covered by a wealth of absorption lines (some P~Cygni)
due to low ionization species (primarily \ion{Fe}{2}). High ionization
resonance lines of \ion{Al}{3}, \ion{Si}{4}, \ion{C}{4}\ and
\ion{N}{5}\ were also identified.
The absorption line spectrum of \etacar\ is very rich, with at least 3 and
possibly more, absorption components/systems seen.  
Viotti \etal\ \cite{VRC89_UV} note, for example, the
presence of absorption extending to $-800$\,\kms\ (which they associate
with the wind), and possibly to as high as -1240\,\kms\ on the
\ion{Si}{4}\ resonance doublet.

More recently the UV spectrum of the central source has been studied
with the GHRS on the HST by Ebbets \etal\ \cite{EWP97_eta_UV}. Their observations showed that the UV
spectrum of  the central source is that of an early B supergiant except
for the additional presence of low ionization wind features, and is quantitatively
very similar to P~Cygni. However close examination of
the low resolution  spectra reveals important differences.
In particular, the spectrum of Eta Carinae is of lower excitation. It
shows a spectrum that is richer in P~Cygni lines, and which generally 
have stronger emission components. Eta Car also has
a significantly larger terminal velocity than P~Cygni. 


As discussed by Hillier \etal\ \cite{HCNF98_HD316} the optical spectrum
of P~Cygni is also similar to that of HDE\,316285, and hence it is
not surprising that the \etacar\ central spectrum shows a
spectrum somewhat similar to P~Cygni. However, both HDE\,316285 and Eta Car are of 
lower excitation than P~Cygni, and their wind densities are higher
(as highlighted by their stronger emission line spectrum).
The presence of wind lines due to low ionization species is
simply a consequence of the dense wind --- as you move out in the
wind the ionization state of the gas decreases.

Ebbets \etal\ \cite{EWP97_eta_UV} infer that there are 2 basic absorption systems ---
one centered near $-500$\,\kms, and the other near $-$1100\,\kms, in broad agreement
with the earlier analysis of Viotti \etal\ \cite{VRC89_UV}. The first system is
probably associated with the stellar wind.
The usefulness of the UV absorption line spectrum for the analysis of the 
central source is unknown.
In other hot stars the UV spectrum is of crucial importance, but because
of intervening circumstellar (e.g., the Little Homunculus and the
Homunculus) and interstellar material it is not easy to 
determine the pure
star+wind spectrum of the central source associated with \etacar.
It is clear that in both optical and UV spectra absorption associated
with the wind is seen, and is characterized by a maximum velocity of
approximately 500\,\kms. In many lines, particularly on the stronger resonance
lines, absorption at higher velocities is also seen. Given the
complex spectrum (see Sect.~\ref{Sec_line_id}) and severe
line blending the determination of accurate line profiles is difficult.

Recently numerous narrow absorption lines, arising in the circumstellar ejecta,
have been found in high resolution HST MAMA spectra
(Gull \etal\ \citeyear{GVB05_abs};
Nielsen \etal\ \citeyear{NGK05}). Two principal systems
have been identified, and are easily recognizable
in the NUV R=110,000 spectra extracted with 0.091\arcsec\ centered on the star. 
The low velocity system has a radial velocity 
of $-146\,$\kms, and is seen primarily in singly ionized species
(e.g. \ion{Fe}{2}, \ion{Ni}{2}, \ion{Cr}{2}).
The populations of the absorbing system are characterized by
a thermal temperature of 6500K across the spectroscopic maximum,
but cool to 5000\,K briefly during the minimum
(Nielsen \etal\ \citeyear{NGK05}, 
Gull, Nielsen \& Vieira Kober \citeyear{GNV06}).
The second system has radial velocity of $-513$\,\kms, similar to
the terminal velocity of the wind. It exhibits lines due to
neutral and singly ionized species (e.g., 
\ion{Fe}{1}, \ion{Fe}{2}, \ion{V}{2}, \ion{Ti}{2}), as well as due to H$_2$.
The observed transitions arise in a gas collisionally populated at 760\,K.
Both absorption systems are characterized
by their narrow line widths which indicate Doppler velocities of
less than 10\,\kms. Using the excitation of the absorption components,
and the observation that the absorption components are not only
seen against the star, it is believed that the absorption systems
arise in the wall of the little Homunculus ($-146$\,\kms\ system;
Gull \etal\ \citeyear{GVB05_abs}; Smith \citeyear{Smi05_lit-hom}) and
in the Homunculus ($-513$\,\kms\ system; Gull \etal\ \citeyear{GVB05_abs}).
The $-513$\,\kms component is consistent with the radial velocity of the H$_2$
emission which cross the line of sight to the star (Smith \citeyear{Smi05_lit-hom}). 
A complete  ejecta spectrum, together with line identifications,
is available electronically (Nielsen \etal\ \citeyear{NGK05};
Gull \etal\ \citeyear{GNV06}).

Weaker narrow absorptions are identified at intermediate
velocities, especially between $-385$ to $-509$\,\kms\ and appear to be
decreasingly excited towards higher velocities. Wind lines of these same
species and indeed these same lines are present in the spectrum. The
difference between ejecta and wind lines is twofold: wind lines, while
lumpy in absorption profile, are continuous with terminal velocities around
520 to 600\,\kms. Strong lines originating from lower levels of Fe II are
dominated by the wind to the point that the ejecta contribution is likely
not to be detectable. 


\section{Identification of Wind Lines}
\label{Sec_line_id}

Due to severe blending, and severe contamination by the
nebular absorption spectrum, line identification in the wind spectrum is difficult, 
and prone to error. The principal culprit for the blending is \ion{Fe}{2} ---
its spectrum dominates almost the entire UV wavelength region.
An obvious solution to the line identification problem is to 
use a model to assist in the identification of lines.
We have done this but, unfortunately, our models
don't provide a perfect fit, and hence there is 
still room for debate about some identifications.

\begin{figure}
\includegraphics[scale=0.7,angle=-90]{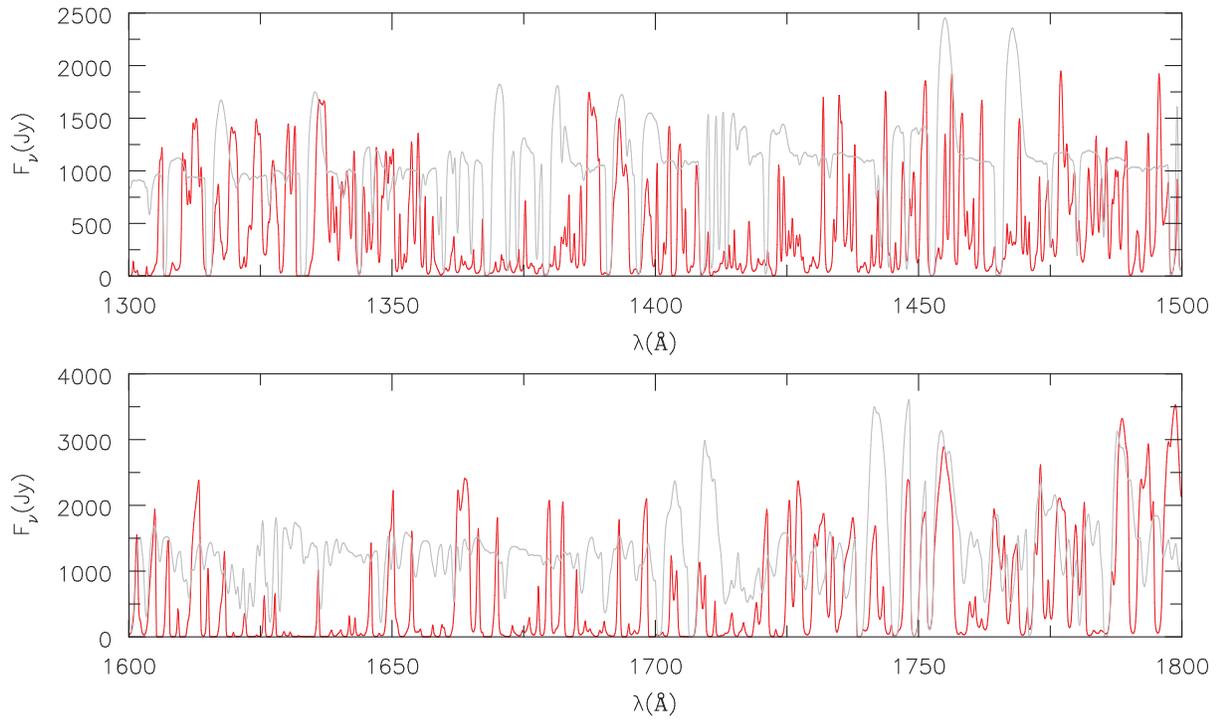}
\caption[]{Illustration of a section of the theoretical UV spectrum
for Eta Carinae. The full spectrum is shown in black [red], while
in gray we show
the spectrum that we obtain when we neglect all bound-bound
transitions except those due to nickel. The nickel spectrum (primarily due to \ion{Ni}{2})
is very rich, however it is masked by \ion{Fe}{2}\ absorption and scattering.}
\label{Fig_nick}
\end{figure}

To highlight the difficulty of analyzing the UV spectrum
we illustrate in Fig.~\ref{Fig_nick} a section of the theoretical UV spectrum. 
Also shown is the same theoretical spectrum 
in which we have omitted the influence of bound-bound transitions
due to all species except nickel. The nickel spectrum is seen to be quite rich,
however in the complete spectrum it is masked by \ion{Fe}{2}.

\begin{figure}
\includegraphics[scale=0.7,angle=-90]{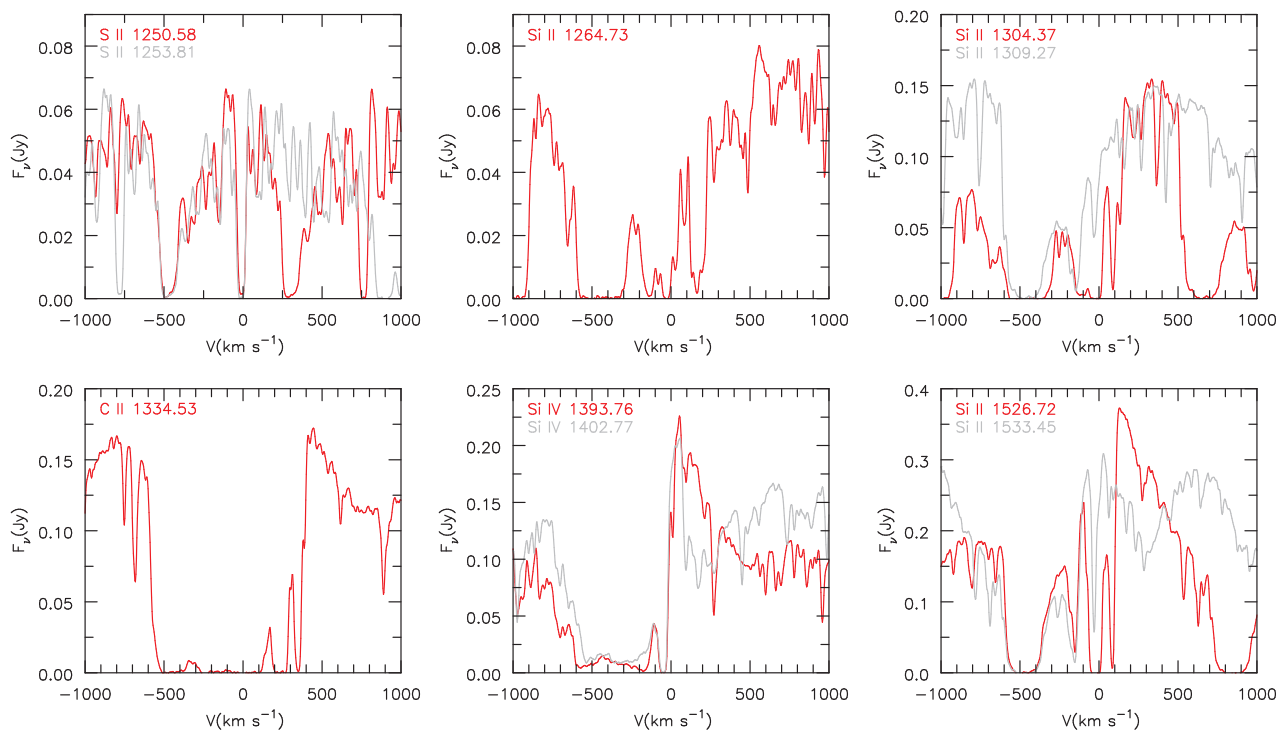}
\caption[]{Selection of strong, and relatively blend free profiles, in
the observed spectrum of \etastar. For some species, another component of
the same multiplet has been overplotted (light gray). 
The steep blue edge of the
profiles indicates a wind terminal velocity greater than 500\,\kms,
but less than 600\,\kms.
Notice the poor agreement between the
two P~Cygni profiles belonging to the \ion{Si}{4} doublet, indicating the importance
of blending with lines due to other species (primarily \ion{Fe}{2}).}
\label{Fig_uvprof1}
\end{figure}

\begin{figure}[h]
\includegraphics[scale=0.7,angle=-90]{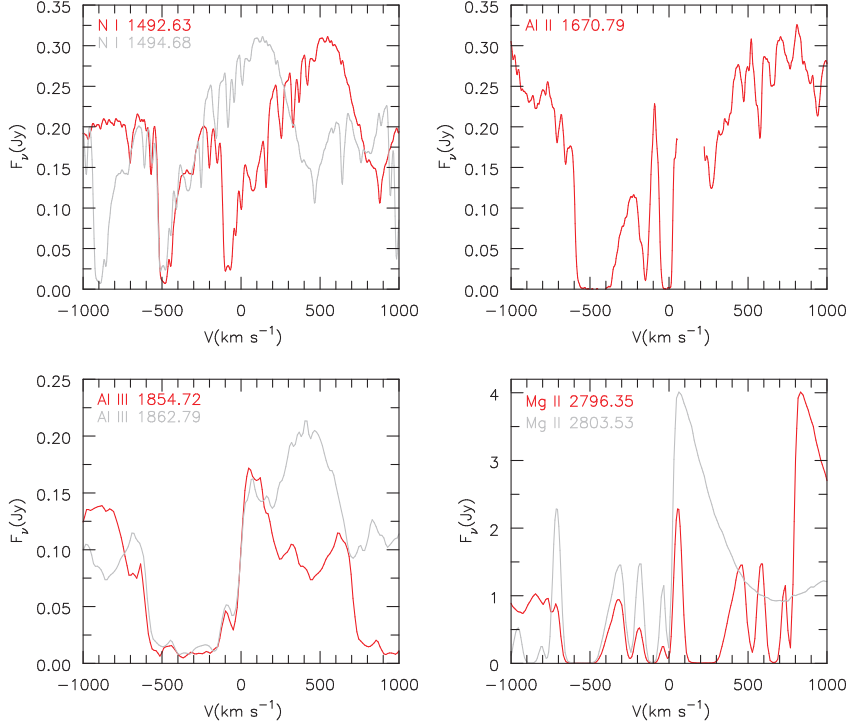}
\caption[]{As for Fig.~\ref{Fig_uvprof1}. The \ion{N}{1} 
profiles are somewhat different to the other lines shown.
The emission components are very obvious, and the absorption
troughs are detached. The later is probably a consequence of the
ionization structure of the wind --- in the inner wind N is
ionized. In this regard the absorption profiles are more
similar to those seen on \ion{Fe}{2}\ lines
(see Fig.~\ref{Fig_fe2_a}).}
\label{Fig_uvprof2}
\end{figure}

\begin{figure}
\includegraphics[scale=0.7,angle=-90]{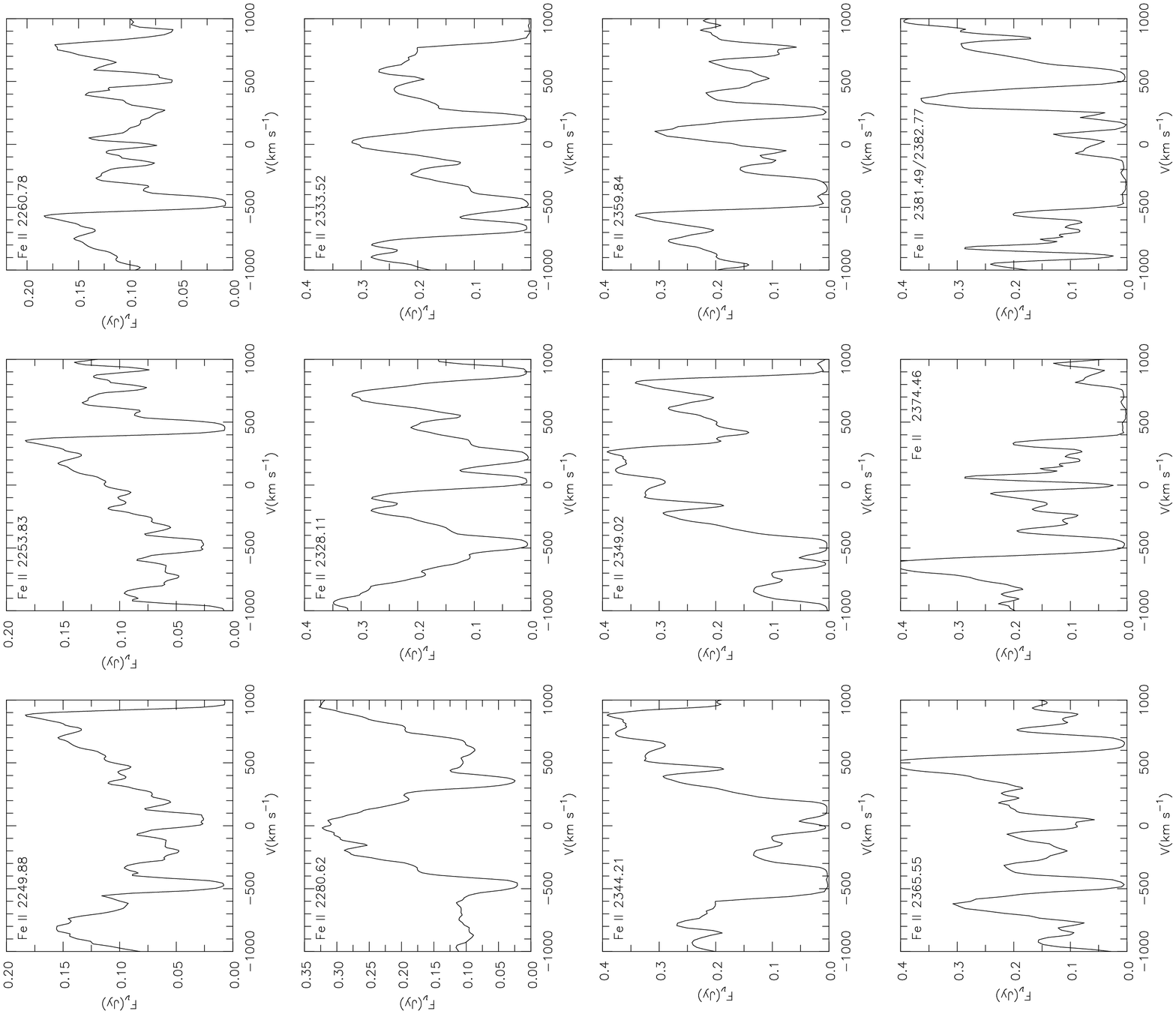}
\caption[]{Selection of \ion{Fe}{2}\ P~Cygni profiles.
These profiles typically show broad absorption between
$-400$ and $-500$\,\kms, again indicating that the terminal velocity
of \etastar's wind is around 500\,\kms. The characteristic
detached absorption associated with the \ion{Fe}{2}\
line profiles arises since Fe is predominantly
Fe$^{2+}$ in the inner wind, but recombines to
Fe$^+$ in the outer wind (see HDIG for further details).
The profiles of these lines might be influenced by ejecta ( 
Nielsen \etal\ \citeyear{NGK05}), although the terminal
velocities are similar to that seen in other UV lines.}
\label{Fig_fe2_a}
\end{figure}

An additional constraint on line identification can be made using
multiplets --- if all (or at least the stronger members) of a multiplet
are seen, and the components have similar profiles, we can be reasonably
assured that the line identifications are correct.
Some of the strongest lines that can be readily identified include \ion{C}{2}
$\lambda$ 1335 (UV1) ,
\ion{Si}{2}  $\lambda\lambda$1304, 1309 (UV3); $\lambda$ 1264 (UV4), 
$\lambda\lambda$ 1527, 1533 (UV 2); $\lambda\lambda$  1808, 1817 (UV1),
\ion{S}{2} $\lambda\lambda$ 1250, 1253 (UV1), 
\ion{Al}{3} $\lambda\lambda$1855, 1863,  \ion{Al}{2} $\lambda$1671, 
\ion{N}{1}\ $\lambda\lambda$ 1493, 1495 (UV4), 
\ion{Mg}{2} $\lambda\lambda$ 2796, 2803, as well as \ion{Fe}{2} lines,
too numerous to list. Some of these lines are illustrated in
Figures \ref{Fig_uvprof1}, \ref{Fig_uvprof2} and \ref{Fig_fe2_a}. 
Particularly striking
is that all lines indicate a terminal velocity for Eta's wind
of between 500 and 600\,\kms\ (Sect.~\ref{Sec_vinf}). Interestingly,
blue shifted absorption ($\approx -400$ to $-500\,$\kms) 
due to \ion{O}{1}\ $\lambda\lambda$ 1302,1304,1306 also appears to be present.

\begin{figure}
\includegraphics[scale=0.7,angle=-90]{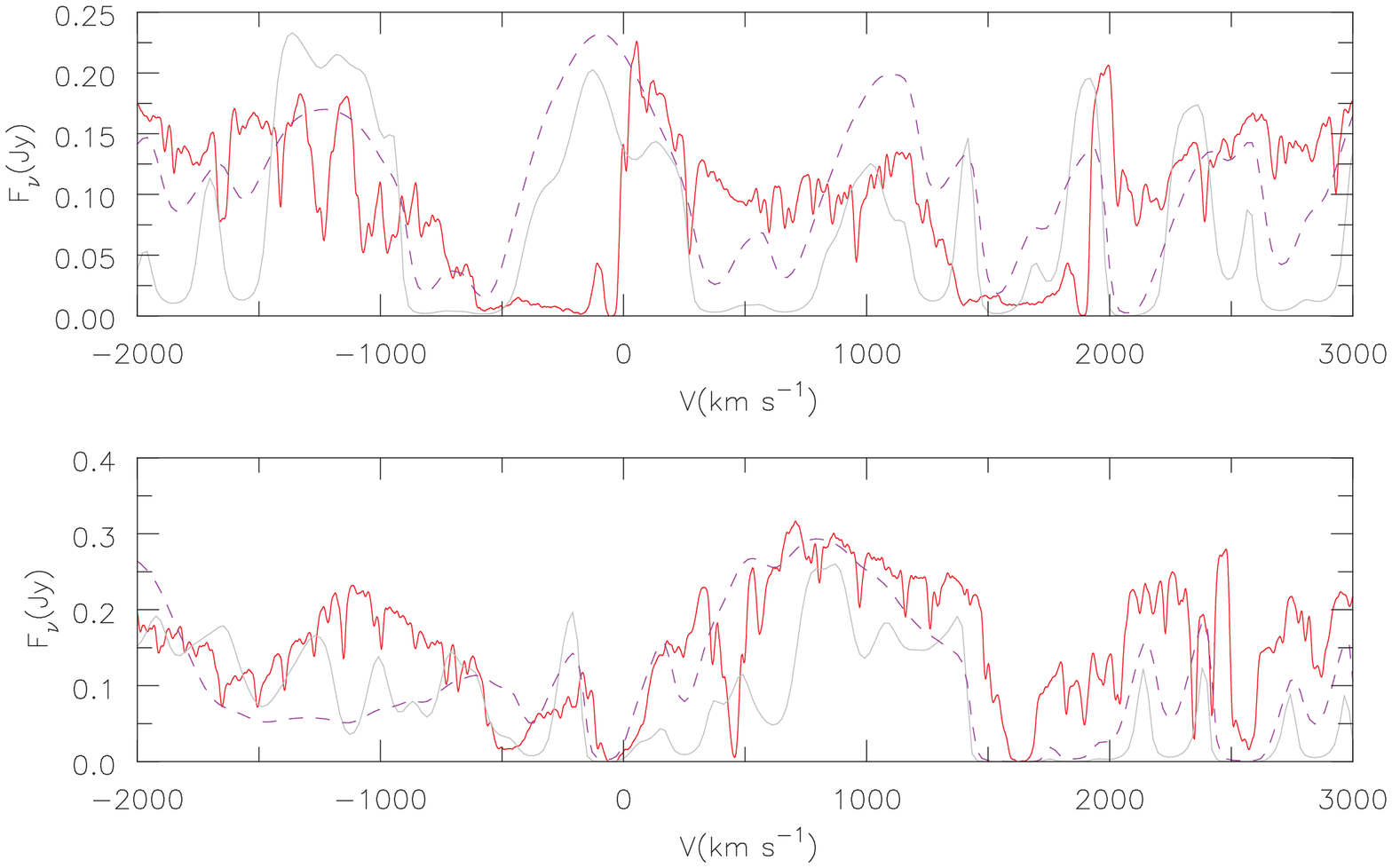}
\caption[]{Illustration of the spectrum of Eta Carinae around the
\ion{Si}{4}\ doublet (top) and the \ion{C}{4}\ doublet (bottom).
The velocity scale is for the blue component of the doublet. For \ion{Si}{4}\
the red component is shifted by approximately 1940\,\kms, and for \ion{C}{4}\
the velocity shift is approximately 500\,\kms. In each plot the solid dark
(red) curve shows the spectrum of Eta Car in 2002-July-04, the full gray curve
shows the full model spectrum,
while the dashed (purple) curve shows the spectrum originating outside
0.033\arcsec\ (see Sect.~\ref{Sec_UV_mystery}). There is
no contribution by \ion{Si}{4}\ or \ion{C}{4}\ to the theoretical
spectrum.}
\label{Fig_sil_carb}
\end{figure}

\begin{figure}
\includegraphics[scale=0.7,angle=-90]{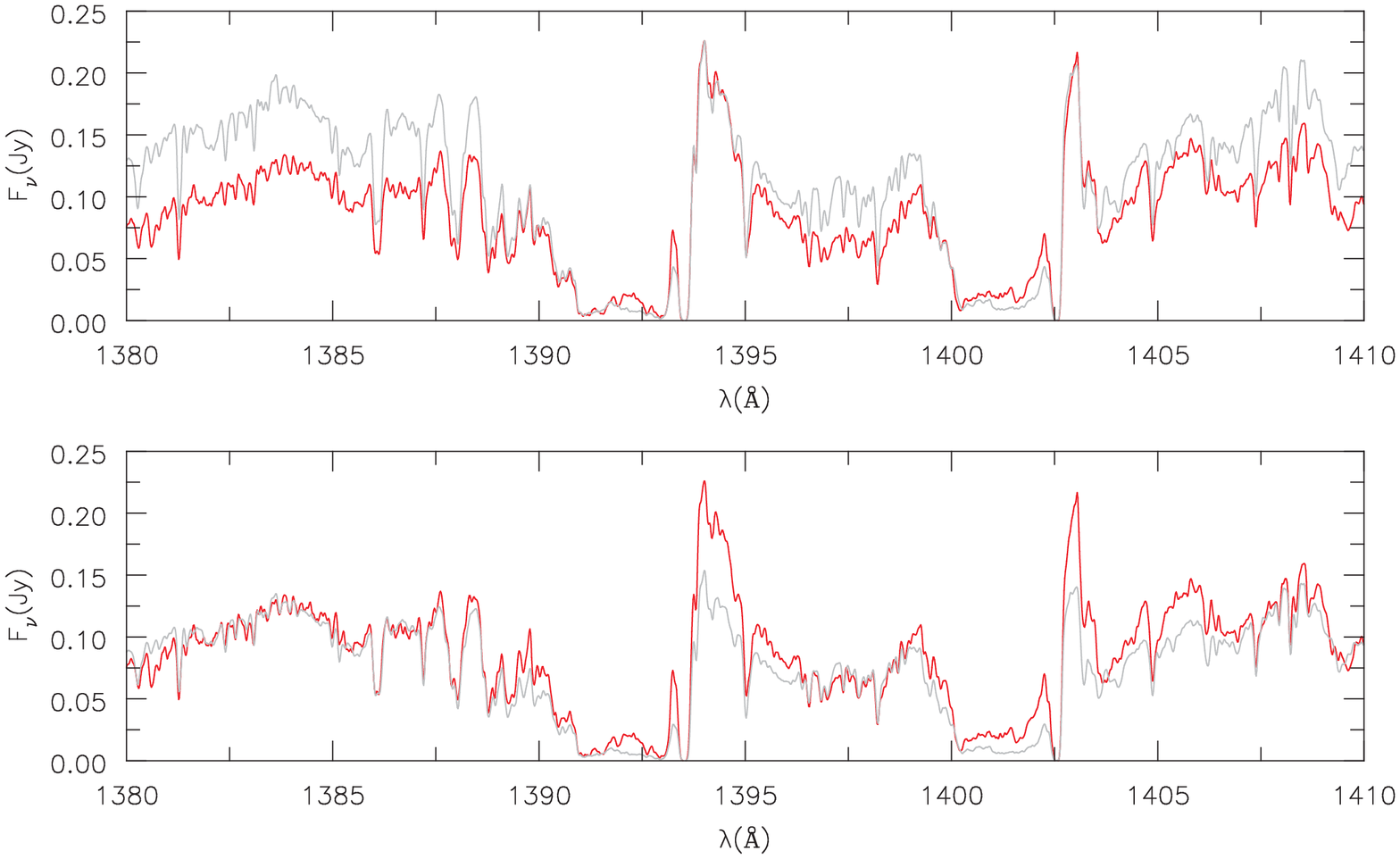}
\caption[]{Illustration of the spectrum of Eta Carinae around the
\ion{Si}{4}\ doublet for the data sets from 23 March, 2000
(black [red]) and 4 July 2002 (gray). In the top plot absolute
fluxes are compared, while in the bottom plot the
latter data has been scaled so that the continua
match. Notice how the top panel suggests that emission
in the \ion{Si}{4}\ lines has not changed with time.
A direct comparison of the profiles of the two \ion{Si}{4}\ 
components is shown in Fig.~\ref{Fig_uvprof1}).}
\label{Fig_sil_var}
\end{figure}

\ion{Si}{4} $\lambda\lambda$1394, 1403 
has often been identified in the
spectrum of Eta Car (e.g., Viotti \etal\ \citeyear{VRC89_UV}; 
Ebbets \etal\ \citeyear{EWP97_eta_UV}). Its presence would not be
surprising, since it is clearly seen in the spectrum
of P~Cygni, however as noted earlier, the spectrum of P~Cygni is generally of
higher excitation. A closer examination of the spectrum and
our theoretical models,
however, reveals some difficulties with the identification.
First, the emission and absorption components of
one component don't perfectly match that of the other
component (Fig.~\ref{Fig_uvprof1}). This could indicate
a mis-identification, or could, not surprisingly, be simply
due to the effect of blends.
Second, the overall excitation of the spectrum is
generally lower than P Cygni. Third, while our
theoretical spectrum apparently shows the apparent presence
of the \ion{Si}{4} lines, they are not actually due to \ion{Si}{4} ---
rather they are a blend of \ion{Fe}{2}\ features.
Because the theoretical model is not a good fit, it is hard to
draw a firm conclusion. The complex spectral region around
the \ion{Si}{4}\ doublet is shown in Fig.~\ref{Fig_sil_carb}.
Notice the complexity of the region, and how apparently
isolated features are severely affected by line blending.
An argument in support of the \ion{Si}{4}\ identification
comes from comparing data sets obtained in March 2000 with
those obtained in July 2002. The two components show similar 
variability behavior.
Interestingly, the simplest interpretation of the variability
is that the \ion{Si}{4}\ profiles have remained the same, while 
the surrounding spectrum has changed
(Fig.~\ref{Fig_sil_var}).

We also find it difficult to firmly identify \ion{C}{4}\ $\lambda
\lambda$ 1548, 1551 in the spectrum. As noted earlier,
\ion{C}{4} has been identified in the spectrum of Eta Carinae
(e.g., Viotti \etal\ \citeyear{VRC89_UV}; 
Ebbets \etal\ \citeyear{EWP97_eta_UV})
While two P~Cygni absorption
components are seen to occur at approximately the right wavelengths,
the features can also be attributed to \ion{Fe}{2} (Fig.~\ref{Fig_sil_carb}).

In pure theoretical models \ion{Si}{4} and \ion{C}{4} do
not substantially influence the spectrum of Eta Carinae. If the 
observed absorption features are due to \ion{Si}{4}\ and \ion{C}{4}\
it may be possible to reconcile the model with them by 
using X-rays. Eta Carinae is a significant source of
X-rays, assumed to arise from the wind-wind interaction,
and these could enhance the Si$^{3+}$ and C$^{3+}$
populations through Auger ionization, especially
since Si$^+$ and C$^+$ are the dominant ionization stages in
the outer wind (Note: Auger ionization typically ejects 2 electrons). 
Auger ionization, for example,
is used to explain the anomalous ionization seen in
O and B stars (\eg\ Cassinelli \& Olson \citeyear{CO79_xray},
MacFarlane \etal\ \citeyear{MWC93_xrays}, 
Pauldrach \etal\ \citeyear{PKP94_zpup}). In single OB stars the
X-rays are generally thought to arise from radiative driven
shocks in the wind
(\eg\ Lucy \& White, \citeyear{LW80_shock}, 
Owocki, Castor \& Rybick \citeyear{OCR88}),
although magnetic fields may also play a role (Babel \& Montmerle \citeyear{BM97_mag};
ud-Doula \& Owocki \citeyear{02mag_field}).
Because the terminal velocity of \etastar's wind is relatively slow
compared to O stars,
radiative driven wind shocks are probably not an important source
of X-rays.
If Auger ionization is important, we might expect to
see significant variations of the high-excitation lines during the X-ray minimum.
Alternatively, ionization of Eta Carinae's wind by UV flux from
the companion star could also produce significant Si$^{3+}$ and C$^{3+}$,
and this would also vary with orbital phase.

Our model (see Sect.~\ref{Sec_model}) is able to predict the presence of most
of the lines shown in Figures \ref{Fig_uvprof1}, \ref{Fig_uvprof2},
and \ref{Fig_fe2_a}. The exceptions are the higher excitation lines
such as those of \ion{Si}{4}\ and \ion{Al}{3} which, as noted
earlier, may be explained by invoking the influence of X-rays, and/or
the ionizing field of the companion, on the wind of the primary.

\section{The Terminal Velocity of the Wind}
\label{Sec_vinf}

Of fundamental importance to understanding the wind
dynamics is the terminal velocity, \Vinf, of the stellar wind.
The terminal velocity can also be used, through the theory
of line driven winds (e.g., Kudritzki \& Puls \citeyear{KP00_rev};
Kudritzki \citeyear{KHP92_rad_driv_wind}) to determine the effective escape
velocity from the ``surface'' of the star.
The terminal velocity, along our line of sight,
 is most easily measured from the blue edge of strong 
UV P~Cygni profiles. While the severe blending
in the UV spectrum of Eta Car makes this difficult, 
it is possible to deduce a value for \Vinf\
using a combination of lines, especially doublets. 
Using the steep blue edge of the absorption profile,
we derive estimates of the terminal velocity along our sight line that 
range from a low of 480\,\kms\ to a high of
around 580\,\kms\ (see Figures \ref{Fig_uvprof1} \& \ref{Fig_uvprof2}). 
The \ion{Fe}{2}\ absorption P~Cygni profiles also give terminal
velocity between 500 and 600\,\kms
(see Fig. \ref{Fig_fe2_a}). We suggest that the true terminal 
velocity is around 500\,\kms, and that the higher derived values are 
probably the result of turbulence in the wind.


As noted previously, it is possible that the wind of
Eta Carinae is axisymmetric, and thus the terminal
velocity may have a latitude dependence.
The angle of our sight line to the bipolar axis of the
Homunculus is approximately $40\,\deg$ (e.g., 
Allen \& Hillier \citeyear{AH93_hom}, 
Davidson \etal\ \citeyear{DSG01_hom_shape}, 
Smith \citeyear{Smi02_IR_Image}).

In earlier UV analyses high-velocity absorption
components extending to 
800\,\kms, 1240\,\kms\ (Viotti \etal\ 1989) and
1100\,\kms\ (Ebbets \etal\ 1997) have been identified.
We find it difficult to confirm these identifications.
Sometimes a feature is seen to be associated with one component,
but not with the second. Narrow ejecta components have been identified
at approximately 1150 and 1175\,\kms (in \ion{Al}{2},
\ion{Al}{3}, \ion{Si}{2}, \ion{Si}{4}, and \ion{Mg}{2}),
and at 1650\,\kms (in \ion{Al}{3}, \ion{C}{4}, and \ion{Si}{4})
(Nielsen \etal\ \citeyear{NGK05}).

How can we reconcile the apparent absence of the high velocity
(broad) absorption? First, we note that many narrow
components arising from other species are present, and
these can sometimes lead to a mistaken identification,
especially with lower spectral resolving powers. These
narrow components are much more readily identified in
our high resolution MAMA spectra. Second,
our models reveal how severe the line blending is in Eta Car, making
it more difficult to be confident in line identifications. Third,
IUE spectra were recorded through large apertures
($10\arcsec\times 18\arcsec$ or 3\arcsec). They thus have a larger contribution to the
spectrum from scattered light (arising from the Homunculus and outer
wind). We note, for example, that Smith
\etal\ \cite{SDG03_eta_lat} identify velocities approaching
1000\,\kms\ from the P~Cygni absorption seen in reflected
spectra of the star taken along the polar axis. Intrinsic
variability may also be important --- velocities
approaching 1000\,\kms\ were seen in some 
H$\alpha$ HST STIS observations of the central star 
(Davidson \etal\ \citeyear{DMH05_Ha}).

\section{Influence of the Companion Star}
\label{Sec_companion_star}

It is now commonly accepted that Eta Carinae is a binary star. However,
no direct influence of the companion on the observed optical or
UV spectrum of \etastar\ has been
seen, although Iping \etal\ (2005) argue the FUSE spectrum
is dominated by the companion star (see \ref{Sec_fuse}).
Recently Steiner \& Damineli \cite{SD04_4686} 
detected broad \ion{He}{2}\ $\lambda 4686$
emission in the stellar spectrum, which might be related to the companion star. 
An extensive discussion of the difficulty in producing this emission
is given by Martin \etal\ \cite{MDH05_4686}, who argue for an alternative
model in which the \ion{He}{2}\ emission arises in a mass ejection. 
Smith \etal\ \cite{SMC04_UV2} previously detected, in the UV scattering
halo, moving shadows which suggest the presence of a companion star. 
Using our model we can ask whether the proposed companion could be observed.

As noted in the introduction, the parameters of the companion are uncertain. 
Analysis of
X-ray data indicates that it has a mass loss rate of approximately 
$\Mdot=1 \times 10^{-5}$\,\Msunyr\ and a terminal velocity
of 3000\,\kms\ (Pittard \& Corcoran \citeyear{PC03_xray}). 
Based on its influence on the spectra of Weigelt blobs,
Verner \etal\ \cite{VGB02_bd_blob} and Verner,
Bruhweiler \& Gull \cite{VBG05_bd_blob}
suggest that the companion is an O type star
with an effective temperature between 34,000\,K and 38,000\,K.
We have therefore adopted the following parameters for the companion:
$\Teff=33,270$\,K, $L= 1.0 \times 10^{6}$\,\Lsun,
$\Mdot=1.0 \times 10^{-5}$\,\Msunyr, $R=30.2$\,\Rsun, and
$\Vinf=3000$\,\kms. The discussion is not significantly influenced by
the choice of these parameters.\footnote{The adopted O star
luminosity is 20\% of that adopted for the model of
the primary star associated with Eta Carinae. Since the
luminosity of the Eta Carinae system is fixed, we should reduce the
corresponding luminosity of the primary. This only adds additional
complications, and does not significantly affect the conclusions.}
The O star companion luminosity is high, and might be expected to have a significant
influence on the circumstellar gas in the neighborhood of Eta Carinae.
A lower luminosity would weaken the influence, and would
also make the star more difficult to detect.

\begin{figure}
\includegraphics[scale=0.7,angle=-90]{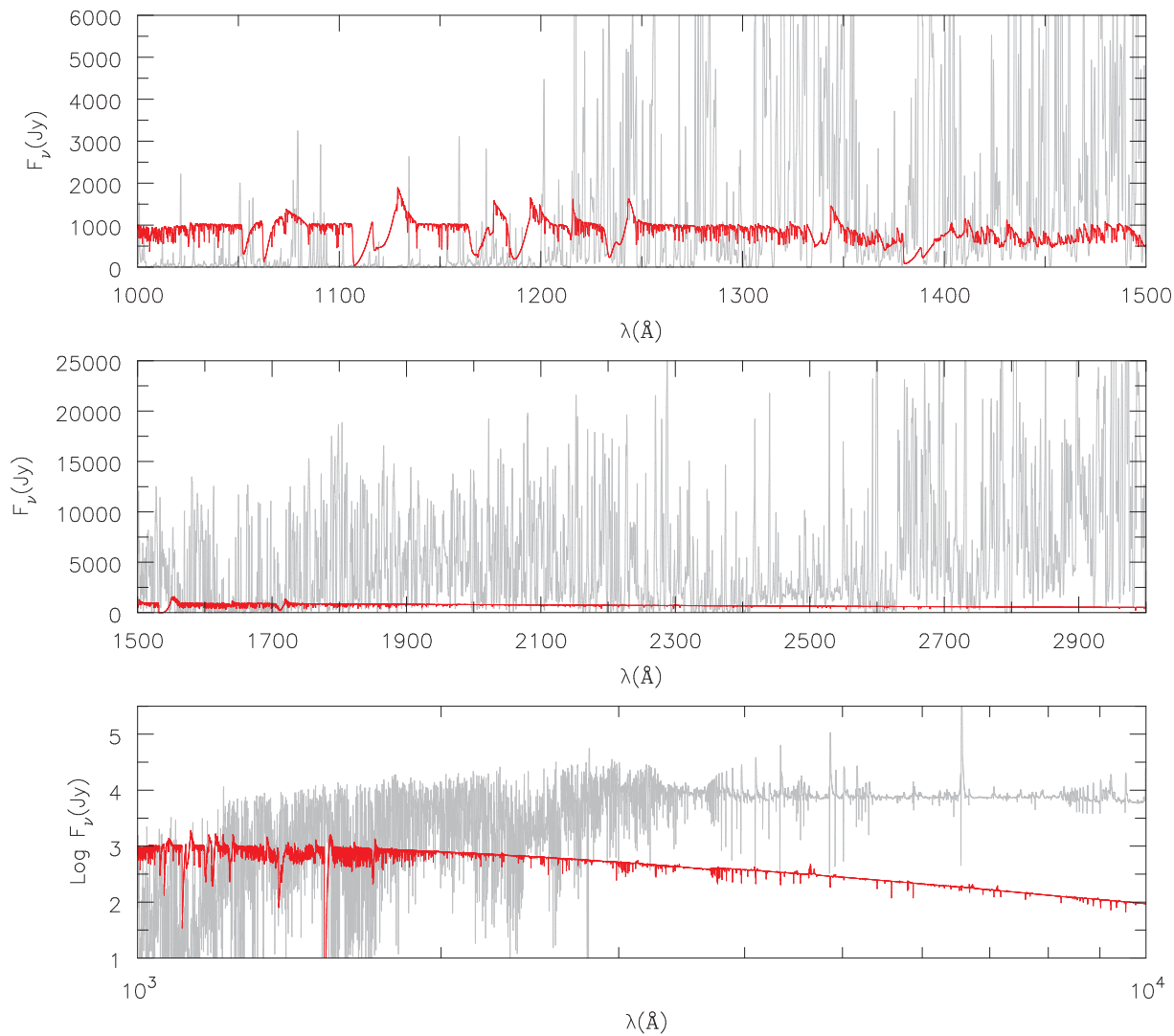}
\caption[]{Illustration of the effect of a companion on the
observed spectra of the central source. The gray line shows
the model spectrum for \etastar, while the dark [red]
line shows the spectrum of the companion star. For the companion 
spectrum, we have not allowed for any absorption by the wind
of the primary.  The spectrum of the companion
star can only be potentially seen shortward of 1500\AA, and is best
observed in the FUSE spectral region.}
\label{Fig_comp_lin}
\end{figure}

In Fig.~\ref{Fig_comp_lin} we illustrate the spectral energy
distribution of the companion and the primary star. As readily
apparent, the primary star dominates the observed spectral energy
distribution, except at wavelengths in the far UV (shortward of
1200\AA). This dominance occurs for three reasons: 
\begin{enumerate}
\item 
  The primary has a higher luminosity. 
\item 
  The secondary is hotter, so more of its energy is emitted in the UV and EUV. 
\item 
  The strong wind of the primary redistributes its UV energy to optical
  wavelengths. 
\end{enumerate}

Thus it is apparent that we would not expect to directly observe the companion
star at optical wavelengths, and that the FUSE spectral range offers the
best wavelength regime to observe the companion star. Indeed a simple
interpretation of the above figures suggests that the FUSE spectrum
could be primarily that of the companion. Could this be the case?

A detailed examination of the FUSE spectral region does not reveal an
obvious O-type star stellar spectrum. However, this is to be expected since
the spectrum of the O star, at most phases, will be strongly modified,
and possibly reprocessed, by the wind of the primary.
In Section \ref{Sec_UV_mystery} we argue that, due to occultation
by dust, the observed UV continuum from Eta Carinae originates at large radii. 
Since the semi-major axis of the O star orbit is approximately 
16\,AU (0.007\arcsec), the O star will generally also be
occulted by the dust.  Consequently its light, to be observable,
must also be scattered into our line of sight.

A more direct consequence of the companion is that it will produce a
significant flux of ionizing  photons. This flux of ionizing photons
will significantly influence the wind of the primary. Indeed, HDIG
suggested that the companion could be responsible for the absence
of strong P~Cygni profiles on the H and \ion{Fe}{2}\ optical lines at most
orbital phases. In the outer wind of Eta, H becomes neutral, and
Fe$^+$ is the dominant ionization stage of iron. Recent HST/STIS
data on the variability of the \ion{He}{1}\ profiles suggest that they
are significantly influenced by the companion star --- indeed 
the direct contribution of the primary 
radiation field to the strength of these lines may be relatively small.

The secondary star, being an O star, emits enough FUV photons to ionize a significant
portion of the primary wind. For the model above, we have that
approximately 1/3 of the luminosity is emitted in the H-Lyman
continuum [giving $\log N(912) =49.7$], while approximately 5\% of the
flux is emitted below 504\AA\ [giving $\log N(504)$=48.6]. The size of
the cavity is difficult to determine since it will be strongly affected by
the shape and density of the wind-wind interaction region, and the
effective temperature of the companion.

In the following analyses we only model the spectrum of the primary
star, and ignore in these models any possible influence of the companion.

\section{The FUSE Spectrum}
\label{Sec_fuse}

The FUSE spectral region is of special interest because potentially
we could directly detect the presence of the companion star
(Sect.~\ref{Sec_companion_star}). Interpretation of the
FUSE spectra, however, is difficult because of the very rich
circumstellar and interstellar spectral features superimposed on the
stellar spectrum, and because of the large aperture that was used. 
It is further confused by the contamination of LWRS
spectra by two nearby field stars (Iping et al. 2005).

The original analysis was carried out using observations obtained
with the LWRS aperture in  2002 June  and 2003 June.
However, as noted earlier, two  B-type stars contribute
approximately half of the observed flux in these spectra. The spectra,
while showing many similarities, do show significant differences
to the narrow aperture observation taken in 2004 April. Almost all of 
these differences
are due to the lack of contamination by the B stars (Iping et al. 
2005). Intrinsic changes in the FUV spectrum of $\eta$ Carinae appear 
to be a minor factor in the differences between the HIRS spectrum and 
earlier LWRS spectra.

The FUSE spectrum overlaps the STIS/MAMA echelle spectrum from approximately
1145 to 1190\,\AA. In this region the MAMA spectrum (2004-Mar-06) is noisy,
and its flux is approximately a factor of 4 below that of the
throughput-corrected  April 10, 2004  FUSE HIRS spectrum ($F_{FUV}\sim 1\times 10^{-12}$ \,erg 
cm$^{-2}$ s$^{-1}$ \AA$^{-1}$).
While some of this may be due to changes in flux during the
month spacing, most is due to the larger FUSE aperture
The FUSE and MAMA spectra
are qualitatively similar, although some features are different.

\begin{figure}
\includegraphics[scale=0.7,angle=-90]{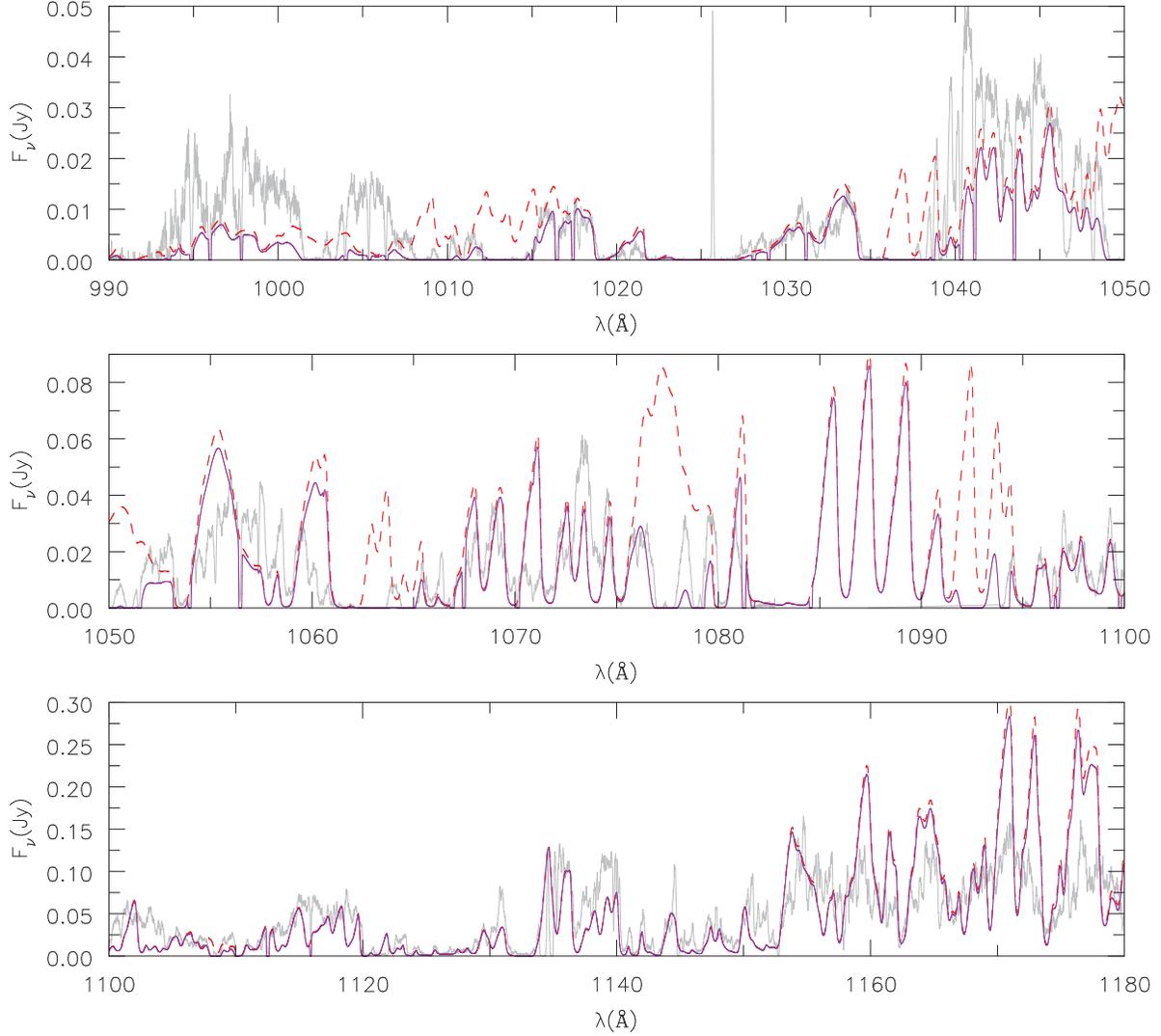}
\caption[]{Comparison of the model spectrum for \etastar\ 
originating outside 0.033\arcsec\ with the FUSE spectrum (gray). Both model spectra
were reddened using E(B-V)=0.4 and R=3.1, and scaled by a factor of 0.5
so as to better fit the observations. The adopted H column density was
$\log \, N(H)=21.8$. To show the importance of molecular hydrogen,
we have provided two plots -- one with negligible H$_2$
absorption (dashed red line), and one with $\log \, N(H_2)=21$
(solid purple line). A detailed fit of the complex $N(H_2)$ spectrum, which has
multiple velocity components (Iping \etal, in preparation),
is beyond the scope of the paper.
}
\label{Fig_fuse_prim}
\end{figure}

The full model spectrum provides an
extremely poor fit to the FUSE observations. A much better fit,
although far from perfect (Fig.~\ref{Fig_fuse_prim}),
is provided by the spectrum originating
outside 0.033\arcsec\ (see Sect.~\ref{Sec_UV_mystery}).
For the comparison we reddened the model using
the Cardelli \etal\ \cite{CCM88_ext}
extinction law with E(B=V)=0.4 and R=3.1,\footnote{
Using R=5, which might be appropriate, affects the scaling
between the model and the FUSE observations but not the overall fit.} and divided
the flux by another factor of 2. Thus to make the comparison, the
reddening was considerably reduced over that needed to fit the
near UV and optical, and is more similar to the expected interstellar
reddening towards Eta Carinae. As expected, the model fits this FUSE 
observation better than the earlier observations made with the LWRS aperture.

\begin{figure}
\includegraphics[scale=0.7,angle=-90]{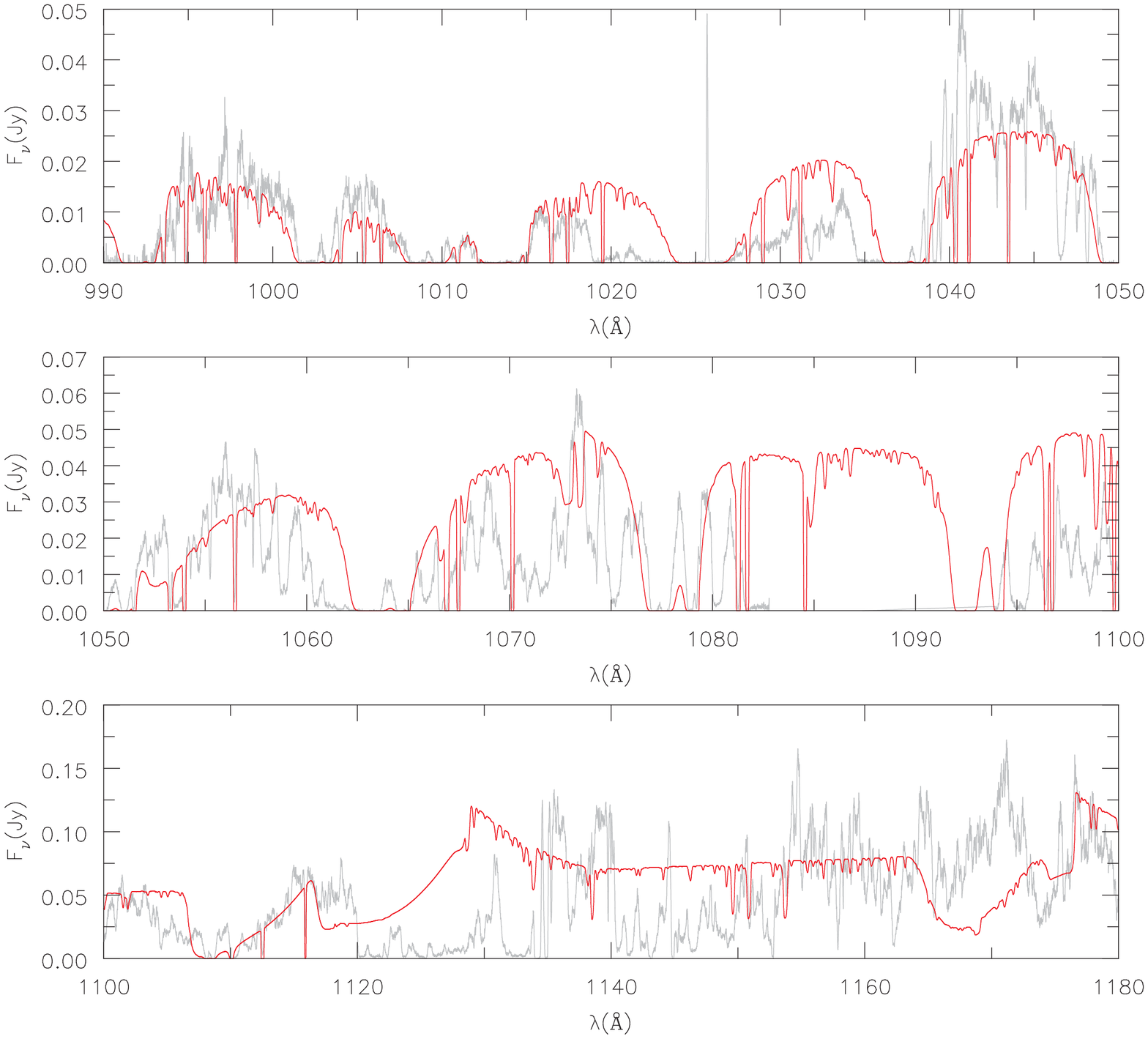}
\caption[]{Comparison of the model spectrum of the companion star
(black [red]) with the FUSE spectrum (gray). The model spectra
were reddened using E(B-V)=0.4 and R=3.1, and then scaled
to better fit the observations. The adopted H column density was
$\log \, N(H)=21.8$, while for H$_2$ is was $\log \, N(H_2)=21.0$.
The influence of H$_2$ can be inferred from the previous figure.}
\label{Fig_fuse_comp}
\end{figure}

Comparison of our theoretical companion spectrum with the FUSE spectrum
does not provide any convincing evidence for standard features 
expected to be present in the FUV spectrum of a companion star.
For example, there are no obvious P~Cygni profiles that could be
unambiguously interpreted as arising in an O-star companion
(see Fig.~\ref{Fig_fuse_comp}). Such P~Cygni profiles should
be prevalent if the O star's mass loss rate, as X-ray studies suggest, is
as high as $10^{-5}$\,\Msunyr. As can be seen from Figures
\ref{Fig_fuse_prim} and \ref{Fig_fuse_comp}, the model
spectrum originating outside 0.033\arcsec\ provides a better fit to the
observations than the companion star. In some sense the invisibility
of the companion star is not surprising --- unless we are looking directly
down the cone caused by interaction of the two stellar winds, the
companion spectrum will be strongly modified by the primary's wind.

What would be the effect of a choice of different parameters for
the companion star? Unless we reduce the luminosity significantly, the
light from the companion is still likely to dominate the FUSE spectral region.
For example, if we increase the effective temperature of the companion
to 40,000\,K the flux in the FUSE spectral region would decrease
by only 25\%. More importantly, the number of \ion{H}{1}\
and \ion{He}{2}\ ionizing photons would increase
(roughly by a factor of 2) meaning that the companion would have
a larger direct influence on the ionization structure of the wind.
A reduction in mass loss by a factor of 5 (and \Vinf\ from 3000 to 2000\,\kms)
reduces the strength of the wind features, but they would still be
easily detectable.


Recently Iping \etal\ (\citeyear{ISG05_Fuse_let}) analyzed FUSE
data of Eta Car taken at several different two epochs. One data set
was obtained close to the 2003 spectroscopic event on 2003-June-27
(X-ray minimum began at 2003 June 29; Corcoran \citeyear{Cor05_xray_mon}). 
The spectrum taken near the spectroscopic minimum has a different 
character to all the
other data sets, showing only a small flux in limited
wavelength regions. Based on several arguments Iping \etal\
concluded that the FUSE flux was primarily due to the companion, and that
during the event the companion's flux was being eclipsed/absorbed
by the primary star and its wind. An alternative explanation
is that the flux in the FUSE spectral region is reduced because
of a shell ejection, as invoked to explain other spectroscopic
features (e.g., Davidson \etal\ \citeyear{DMH05_Ha},
Martin \etal\ \citeyear{MDH05_4686}, Smith \etal\ \citeyear{SDG03_eta_lat}).

The largest uncertainties on the predicted FUSE spectrum,
and on the interpretation of the observed FUSE spectrum,
arise from the affect of the dust curtain, and on how the
light from the companion is modified by the wind of the primary,
an important effect not included in the present analysis. 
The study of these effects is beyond the scope of the present paper.

\subsection{\ion{H}{1}\ Column Density}
\label{Sec_HI_col}

\begin{figure}[tb]
\includegraphics[scale=0.7,angle=-90]{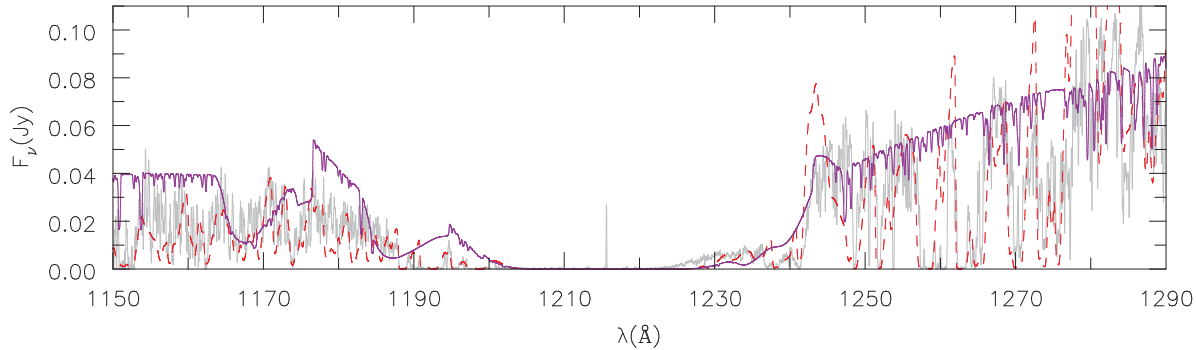}
\caption[]{Illustration of Ly$\alpha$ region in \etastar. 
We show the model spectrum for \etastar\
(dashed red line), and the model of the O star companion (solid purple line) modified
assuming an interstellar H column density of $\log \, $N(\ion{H}{1})$=22.5$.
This column density provides a reasonable fit to the
Ly$\alpha$ profile in the observed spectrum (gray). The normalization
is somewhat arbitrary, since the reddening (and reddening law)
is virtually impossible to determine. 
On the right we see no evidence for \ion{N}{5}\ in the companion
spectrum (since it is cool and X-rays were not included), in the
heavily blanketed model spectrum, or in the equally
blanketed observed spectrum of \etastar.}
\label{Fig_lya}
\end{figure}

In principle, the \ion{H}{1}\ column densities towards Eta can be measured from 
the Ly$\alpha$ (1216\AA) and Ly$\beta$ (1026\AA) lines in the STIS and FUSE spectra.
In Fig.~\ref{Fig_lya} we show the STIS spectrum in the neighborhood of
Ly$\alpha$. 
A firm upper limit to
the neutral hydrogen column density is $\log \, $N(\ion{H}{1})$=22.7$ which 
primarily is set by the presence of a 
significant flux around 1230\,\AA. The best fit column density, using the
spectrum of the primary originating outside 0.033\arcsec, is
$\log $N(\ion{H}{1})$=22.5$, while the best fit obtained using the
spectrum of the secondary is 22.3. The lower limit is somewhat difficult
to determine but $\log $N(\ion{H}{1})$=22.0$ is a reasonable estimate.
Because the spectrum of the primary provides the best fit to the
interstellar/circumstellar Ly$\alpha$ profile 
we have adopted $\log \, $N(\ion{H}{1})$=22.5$. 

Since the dust may not be uniform across the STIS aperture
(Hillier \& Allen \citeyear{HA92_eta}; Morse \etal\ \citeyear{MDB98_im};
Smith \etal\ \citeyear{SMG04_UV1}),
and because of the scattering, the derived log N(\ion{H}{1})$=22.5$ must be
considered a lower value. Such a column density is consistent with that
required to cause the narrow Balmer line absorption which
is seen both on and off the star (Johansson \etal\ \citeyear{JGH05_nar_balmer}).
The variable absorption is centered at $-150$\,\kms\ on the star, but shifts 
to $-45$\,\kms\ off the star.  Due to the strong intrinsic
absorption around Ly$\beta$, and uncertainties in the model, it is difficult
to determine a column density from the last FUSE data set
(Figures \ref{Fig_fuse_prim}, \ref{Fig_fuse_comp}). However, the presence of
flux around 1028\,\AA\ suggests that $\log$ N(\ion{H}{1})$<22.5$,
and closer to 21.8. A lower column density for the FUSE data set is
consistent with the idea that the FUSE flux comes from a more
extended region than the STIS data, and suffers less absorption.


\section{The Model}
\label{Sec_model}

The ``final'' model adopted for modeling the optical spectrum of \etastar,
was discussed extensively by HDIG, and has the following parameters:

\indent     $\log L_\ast / L_\odot =6.7 4$, \\
\indent     $\Rstar(\tau=155,V=0.32\,\kms)=60.0\,\Rsun\ \; (\Teff=35,310\,$ K) \\
\indent     $\Rstar^\prime(\tau=10 ,V=177\,\kms)=99.4\,\Rsun\ \; (\Teff=27,433\, $ K) \\
\indent     $\Rstar^{\prime\prime}(\tau=0.67,V=375\,\kms)=881.0\,\Rsun\ \; (\Teff=9,210\,$ K) \\
\indent     $\Mdot  = 1.0 \times 10^{-3} \,M_{\odot}\,{\rm yr}^{-1}$, \\
\indent     $V_{\infty}  = 500 $\,{\kms}, \\
\indent     H/He$ =5.0$ by number.  \\
\indent     $f(r)=0.1+ 0.9 \exp(-v(r)/v_{cl})$, $v_{cl}=100$\,\kms,
             $r(v_{cl})\sim 1.25$\Rstar

\noindent
The first radius listed corresponds to the inner boundary
of our model, and is only a few percent smaller than the radius of the star 
at the sonic point.  The other 2 radii correspond to locations in the wind,
and highlight the large optical depth of the wind.
$f(r)$ is the volume filling factor which is used to allow for clumping in the
stellar wind, and in the present models it has a minimum value
of 0.1. Clumping reduces the derived mass-loss rates and reduces the
strength of the electron  scattering wings (Hillier \citeyear{Hil91_es},
Hillier and Miller \citeyear{HM99_WC}).
Recent near-IR interferometric observations by 
van Boekel \etal\ (\citeyear{Boek03}) suggest that
the mass-loss rate has the somewhat higher value of $\Mdot=1.6 \pm 0.3 \times 10^{-3}$\,
\Msunyr. This value gives electron scattering wings
which are somewhat too large. Unfortunately the wind asymmetry,
and the possibility that the HST is still not obtaining the
``true'' spectrum of the star (Hillier \etal\ \citeyear{HDIG_hven_eta}),\footnote{
Several earlier publications (e.g, Hillier \& Allen \citeyear{HA92_eta},
Davidson \etal\ \citeyear{DEW95_FOS}, 
Hillier \etal\ \cite{HDIG_hven_eta}; Smith \etal\ \citeyear{SDG03_eta_lat}) 
also noted that the Balmer lines were stronger (i.e., had
a larger EW) on the star than in spectra taken in the lobes. This is surprising
since the lobe spectrum is primarily produced by scattering of light from
the central source. The effect can be seen for H$\beta$ in one of
the spectra shown in Fig.~\ref{Fig_rw_a}.}
makes a more accurate value difficult to determine. The influence of changes
in mass loss rate on the observed UV spectrum are discussed in 
\S~\ref{Sec_UV_mystery}, while further information, particularly for the
optical spectrum, is provided by HDIG.

We also note that the model is not unique. In particular,
the derived mass-loss rate will be
higher if we adopt a larger He/H abundance ratio.
The value chosen is consistent with that found from 
studies of the S-condensation by Davidson \etal\ \cite{DDWG86}, 
but the stellar abundance could be higher,
especially if recent estimates of the mass of the Homunculus
($> 10\,\Msun$; Smith \etal\ \citeyear{SGH03_IR}) are correct. 
Such larger masses would imply that
a significant amount of material was lost from Eta Carinae, potentially
allowing more processed material to be revealed at the stellar surface.
The He/H ratio cannot be derived uniquely from the stellar spectrum
due to degeneracy between the mass-loss rate and the He/H ratio
(HDIG). The derivation of a reliable He/H ratio
is made even more difficult because of the strong
possibility that the \ion{He}{1}\
profiles and fluxes are significantly influenced by the companion star.
 
We also note that, because of the opaque wind, the radius of the star cannot 
be derived.  Models with $\Rstar = $60\,\Rsun\ to 480\,\Rsun\ predict similar spectra,
with the exception that the \ion{He}{1}\ lines weaken as the radius increases.
If the \ion{He}{1}\ lines are primarily produced by the radiation
field of the companion, or if the mass-loss rate is lower than
$10^{-3}\,$\Msunyr, models with larger radii are favored.

Abundances are taken from the analysis of HDIG. In that analysis we
found that the optical observations were consistent with solar mass
fractions for Fe, Si, and Mg. The deduced CNO abundances were consistent
with that expected for full CNO processing, while we found evidence
that the Na abundance was enhanced by at least a factor of 2. Given the
complex UV spectrum, with its often saturated and badly blended
P Cygni profiles, and the complex origin of the UV spectrum (see Sect. 
\ref{Sec_UV_mystery}) we have not attempted to revise these abundances.
This will be attempted in a future study.


 
\subsection{Model Improvements}
\label{Sec_mod_improvements}

The original model for \etastar\ extended to 1000\Rstar (0.1\arcsec).
For this work it was necessary to extend the model out to
20000\,\Rstar (approximately 2.2\arcsec). This is close to the
size, along the polar direction, of the Little Homunculus
(Ishibashi \etal \citeyear{Ish03_lit_hom}).
In order to facilitate this extension we did the following: (a) We
computed models which included adiabatic cooling. (b) We improved the
model \ion{Fe}{2}\ atom so that the lowest energy levels could be treated as
individual super-levels.  This latter change is necessary to model the
optical wind spectrum formed at large radii, and is also important for
model convergence.  Atomic data for our \ion{Fe}{2}\ atom
is from Nahar \cite{Nah95_FeII}; Zhang \& Pradhan \cite{ZP95_FeII_col};
and Kurucz \& Bell \cite{KB95_CD}.
The charge exchange cross-sections were chosen so
the rate was proportional to the statistical weights of the
final level. The rates were scaled so that the total rate between
individual levels was equal to the total rate between the corresponding terms.
Charge exchange rates were obtained from the compilation of Kingdon \& Ferland 
\cite{KB95_CD}.


\section{Asymmetries and Time Scales}
\label{Sec_asym}

The Homunculus shows an obvious bipolar symmetry.  From HST studies
and VLT studies there is also evidence for a wind asymmetry. The wind
seems to be denser, and flow faster, along the polar directions (Smith
\etal\ \citeyear{SDG03_eta_lat}). VLT observations indicate a density contrast of a
factor of 1.5 between the polar and equatorial flows (van Boekel 
\etal\ \citeyear{Boek03}). HST observations
indicate velocities approaching 900\,\kms\ in the polar flow, whereas the
wind in most directions has a velocity of only 500 to 600\,\kms
(Smith \etal\ \citeyear{SDG03_eta_lat}).
It is unclear whether the velocity of 900\,\kms\ corresponds to
the true terminal velocity of the wind. It was measured from the
blue most absorption edge of the H$\alpha$ profile, and is larger than
the value measured from the blue edge of the stronger P~Cygni absorption
trough. Further, velocities measured from emission lines, independent of
orientation, indicate ``mean maximum'' outflow velocities of around 500\,\kms.
In O stars the extended (shallow) absorption is generally thought to arise from shocks in
the stellar wind, and is assumed not to indicate the actual 
terminal velocity of the mean flow. When the minimum of the P Cygni profile
is measured on H$\alpha$ measured velocities, as a function
of latitude, vary from just over 400\,\kms at 45$^\circ$ to
a little under 600\,\kms along the pole (Smith \etal\ \citeyear{SDG03_eta_lat}).
The variation depends on the observational epoch.

In the first modeling of the data we will ignore 
intrinsic asymmetries associated with \etastar's primary wind. 
We will also ignore in the calculations perturbations to the wind structure, and the
ionization state of the wind, induced by the binary companion.\footnote{It 
is important to distinguish between the two type of asymmetries.
An intrinsic asymmetry associated with the underlying
primary has important implications for the mass-loss process, and for
understanding the evolution of the primary star. 
Such an asymmetry may, or may not, be time variable.
In the case of the binary the asymmetry is more complex, because
it involves both the interaction of the two stellar winds, and the influence
of the radiation field of the companion. In addition, the effect of the asymmetry
on the spectrum will change with orbital phase.}
We thus fit the data with some model, which then represents some
``averaged'' wind properties. However, we then utilize a variety of models
to gain insights into the possible influence of deviations from
spherical symmetry. Deviations of the data from the best
model will give insights into the asymmetries and the influence of the
companion. The goal of this work is not to fit the spectrum --- rather 
it is to gain insights into the nature of the central star and its wind. 

One-dimension studies, such as the one performed here, are an absolute 
necessity. Work is in progress to develop 2.5D and 3D radiative 
transfer codes for stars with
extended atmospheres (e.g., Busche \& Hillier \citeyear{BH05_rot},
Georgiev, Hillier \& Zsarg\'o \citeyear{GHZ05_code}, and Zsarg\'o, Hillier \& Georgiev
\citeyear{ZHG05_2D}, van Noort, Hubeny, \& Lanz \citeyear{NHL02}). 
However full non-LTE models, with equivalent complexity (in terms
of the model atoms) require 2 to 3 orders more computational effort.
In addition the parameter space is much larger. Thus 1D models will
still play a crucial role in gaining critical insights, and in limiting the
parameter space to be studied.

We also need to be concerned with variability. Variability implies 
changing physical conditions, and different regions of the flow will respond 
differently, and on different time scales, to these changing
conditions.  First, an estimate of the wind flow time can be written as

\begin{equation}
       t= 21.8 
              \left( {d \over 2.3\,\hbox{kpc} } \right)
              \left( {r \over 1\,\arcsec } \right)
              \left( {500\,\hbox{km\,s}^{-1}\over v } \right) \,\,
\hbox{yrs.}
\end{equation}

\noindent
This time scale is significant: the wind at 1\arcsec\ reflects the
mass loss of \etastar\ 22 years earlier. Consequently the large scale
wind that can currently be directly observed with HST is dependent on
the mass-loss history of $\eta$ Carinae over the last 20 years.
Even at 0.1\arcsec, the flow time is over 2 years. 

We will assume for simplicity, and for lack of other information,
that the flow has been constant. We note, however,
that due to the long flow times the history of the mass loss from
Eta Carinae could be important. Given the erratic variability
exhibited by Eta Carinae over the last 20 years 
(e.g., Whitelock \etal\ \citeyear{WFK94_eta_var},
Sterken \etal\ \citeyear{SGGB99_eta_var}, Martin \& Koppelman \citeyear{MK04_eta_var}),
flow variations could influence the density structure of the
outer wind.

The recombination time-scale is

\begin{equation}
       t=0.32 
          \left( {10^{-12} \over \alpha_{rec} } \right)
          \left( { 10^5  \over N_e } \right) \,\, \hbox{yrs}
\end{equation}
\noindent
For hydrogen, $\alpha_B$ is $2.6 \times 10^{-13}$ at $10^4\,$K 
(Osterbrock \citeyear{Ost_gas_neb}).
In the inner wind the recombination time-scale is much less than the
flow time, but at large radii the recombination time scale can be
longer. In terms of the model parameters it can be written as

  \begin{equation}
      t={34 \over \gamma} \left( {10^{-3} \Msunyr \over \Mdot } \right) \
               \left({f \over 0.1}\right)
               \left({v \over 500\,\kms} \right)
               \left( {10^{-12} \over \alpha_{rec} } \right)
               \left({r \over 1\arcsec}\right)^2 \,\, \hbox{days}
  \end{equation}
where $\gamma$ is the ratio of electrons to atoms. 
In the inner wind $\gamma$ is approximately unity, whereas in the outer wind
$\gamma$ can be substantially less than 1.

\section{The UV Mystery}
\label{Sec_UV_mystery}

In the 2001 paper (HDIG)  our model was unable to explain the UV spectrum. The
predicted UV spectrum (1200 -- 1600\AA) was much richer in UV absorption lines than that
observed. At the time it was noted that a binary companion,
or a wind asymmetry might provide possible solutions to the discrepancy. 
Since that time, direct
evidence for a wind asymmetry has been found 
(Smith \etal\ \citeyear{SDG03_eta_lat}, van Boekel \etal\ \citeyear{Boek03}).
However various observational and theoretical indicators suggest that this
is not the full solution.  
More recently the FUSE spectrum of $\eta$ Carinae has become available. The
FUSE spectrum is also significantly different from the model
predictions. 

\begin{figure}
\includegraphics[scale=0.7,angle=-90]{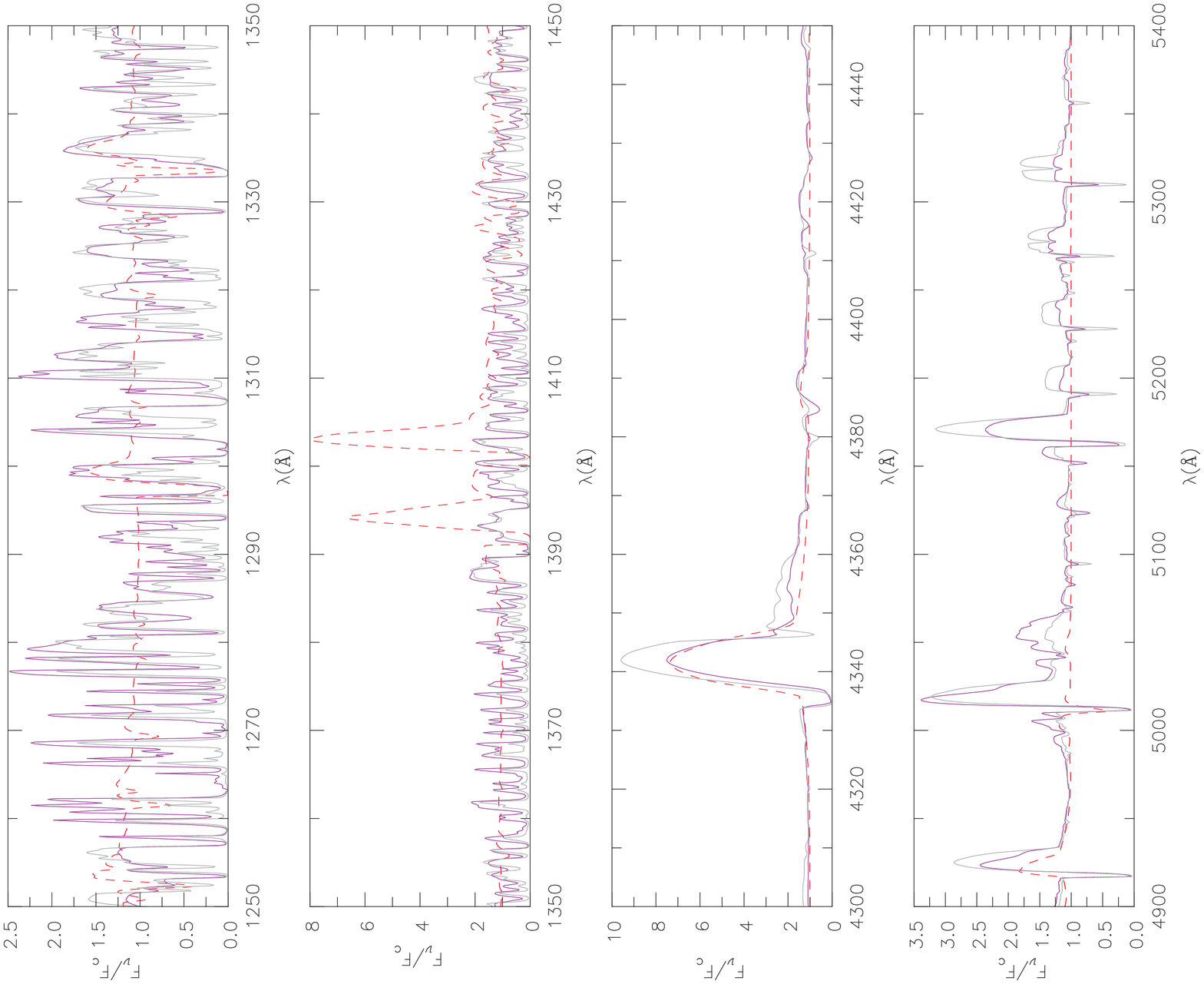}
\caption[]{Illustration of the effect of lowering the mass-loss rate
on the predicted spectrum for \etastar. The 
gray curve is for $\Mdot=1.0\times 10^{-3}$\,\Msunyr,
the purple curve is for $\Mdot=5.0\times 10^{-4}$\,\Msunyr,
and the dashed red curve is for $\Mdot=2.5\times 10^{-4}$\,\Msunyr.
In the lower mass loss rate models \ion{N}{3}\ lines would
become obvious, but in these particular models our \ion{N}{3}\
atoms were limited to the first 8 terms.}
\label{Fig_spec_lowm}
\end{figure}

In order to understand the UV spectrum we investigated a wide range of
models in which we varied the mass-loss rate to see if we could find a
match to the UV spectrum. While the mass-loss rate does have a
significant effect on the UV spectrum, a variation in mass-loss rate alone
cannot account for the observed spectrum. Examples of such spectra are shown in
Fig.~\ref{Fig_spec_lowm}. As the mass-loss rate is lowered, the 
\ion{Fe}{2}\ emission lines weaken, the \ion{He}{1}\ emission lines strengthen,
and the Balmer P Cygni absorption weakens and eventually disappears, as
does much of the \ion{Fe}{2} absorption in the UV. Eventually
\ion{N}{3}\ $\lambda\lambda  4634, 4640/4641$ and \ion{He}{2}\ $\lambda 4686$
come into emission, and neither are observed.
In addition the V flux is reduced: the model with 
$\Mdot=1.0\times 10^{-3}$\,\Msunyr\ has a visual flux 3.3 times larger than the
model with $\Mdot=2.5\times 10^{-4}$\,\Msunyr. Finally we note that in
the low mass-loss rate model Ly$\alpha$ is in emission and contains
approximately 7\% of the emitted flux, while over 20\% of the flux
is emitted shortward of the Lyman jump.

The large differences in the model spectra are related to
the ionization of H and Fe. In the high mass-loss model, H recombines
in the outer envelope.  When the neutral H ionization fraction
exceeds $10^{-4}$, charge exchange processes with Fe$^{2+}$ become
important. This drives the ionization of Fe quickly towards Fe$^{+}$.
In the lower mass-loss rate model, H recombination never
occurs and Fe is primarily Fe$^{2+}$. Smith \etal\ 
\cite{SDG03_eta_lat} argue that this recombination is latitude 
dependent, and that during the event the wind at our latitude,
which is normally ionized, recombines.

It is clear from a comparison of the models with observation that
the low mass-loss rate models do not provide a better fit to the observations
--- indeed the average wind properties appear to be more similar
to our original model. It might be possible to generate a better
fit by using an asymmetric model with a range of mass-loss rates.
In such a model our sight line would preferentially view the
star in a direction of low-mass loss rate, while the \ion{Fe}{2}\
lines would come from another region (presumably a denser polar
flow). Unfortunately such a model, by itself, does not appear to provide an 
explanation for some unexplained UV emission features.
Is there an alternative explanation?

Several observations provide clues to the resolution of the UV mystery.
First, the UV emitting region of $\etastar$ is extended. 
Overwhelming evidence is provided by a comparison of FUSE fluxes with
HST fluxes integrated over the central star. In the overlapping wavebands, 
the FUSE flux (2004-April-11) exceeds the STIS flux (2002-July-04)
by a factor of 4 
(6.7 if we allow for the FUSE point-source throughput-correction).
Similarly, evidence for extended UV emission is 
provided by a comparison of GHRS observations with HST STIS fluxes. As noted by HDIG,
the GHRS fluxes (obtained in 1995) are a factor of 3 higher than those
obtained using STIS. Proof is provided by the 
ACS/HRC images of Eta Carinae, which clearly show
extended UV emission around the star (Smith \etal\ \citeyear{SMG04_UV1}).
The inferred extension of the UV emitting region is not surprising: 
the primary star is only a relatively small direct contributor to the 
flux of $\eta$ Carinae at optical wavelengths.

Several possible mechanisms could provide an explanation for the
extended UV emission: electron scattering, dust scattering, and
resonance line scattering. 

Electron scattering can be ruled out since the expected column
densities are too low. Further electron scattering is wavelength
independent, and this does not agree with the observations.  In the
neighborhood of the star (e.g., inside 0.3\arcsec) the UV emission is more
extended than the optical emission.

The role of dust is more difficult to determine. From infrared
observations we know that dust exists in the walls of the Homunculus,
in the equatorial disk, and in an inner core 
(e.g., Smith \& Gehrz \citeyear{SG98_prop}; 
Smith \citeyear{Smi02_IR_Image};
Smith \etal\ \citeyear{SDG03_eta_lat}).
Further we know
that the central source suffers circumstellar extinction. This
circumstellar extinction is not uniform: the Weigelt B, C and D
blobs, located less than 0.2\arcsec\ from the central source suffer
much less extinction (Hillier \& Allen \citeyear{HA92_eta};
Davidson \etal\ \citeyear{DEW95_FOS}).
If they did not suffer less extinction, we could not explain 
their energetics (Davidson \& Humphreys \citeyear{DH86_blob},
Weigelt \etal\ \citeyear{WAB95_FOC}).
The complex distribution of dust around the central source,
and the complicated scattering geometry, can be seen in the
UV images of Smith \etal\ \cite{SMG04_UV1}. 

The dust sublimation radius is believed to be approximately
0.07\arcsec, assuming a dust sublimation temperature of 1000\,K
(e.g., Smith \etal\ \citeyear{SGH03_IR}).
This radius is similar to the spatial resolution of our HST
observations. Recently Chesneau \etal\ \cite{Ches05_eta_IR}
found indications, from high spatial resolution IR images, that the dust
around \etastar\ seemed to occur outside a radius of
0.130 to 0.170\arcsec.

Dust scattering is very important for generating the Homunculus
spectrum. However various arguments suggest that it is not the {\em
dominant} mechanism close to the star.  First, dust scattering
preserves the basic underlying spectrum. However no theoretical model
spectrum generated to date has been able to match the UV spectrum.
Another potential problem is that dust scattering varies continuously
but smoothly with wavelength. UV observations suggest that the extended UV
spectrum varies strongly with wavelength. This variation is especially
obvious in UV spectra taken during the 2003 event.

These, and other considerations, lead us to believe 
(Hillier \etal\ \citeyear{HDG03_scat}) that the UV
emitting region is very extended, and the UV spectrum we observe
arises from bound-bound scattering in material at large radii (e.g., at
radii $> 0.01$\arcsec).
The UV radiation coming from such radii
suffers less extinction than does the central star, and hence is
more readily detected. Thus while dust does not provide a direct
explanation for the UV halo, it does provide the crucial coronagraph
which preferentially blocks our line of sight to the central star.

It is worth noting that there are two distinct, but related,
processes occurring. First, significant UV emission is coming from
around the ``point source''. This emission is clearly identified in the
images of Smith \etal\ \cite{SMC04_UV2}, and as Smith \etal\ note,
the contribution of the point source to the total light is lower
at UV wavelengths (see below). The visibility of this extended UV emission
is enhanced by dust obscuration of the central source. Second, dust is
obscuring {\em some} of the star and its wind. Because the obscuration
is not uniform, we cannot directly compare observed stellar spectra with
models. Indeed, it is the non-uniform obscuration that allows the broad
[\ion{Fe}{2}] lines to be seen in ground-based optical spectra.

For this model to work the dust must have an asymmetric distribution. This is
in accord with the observations, since observations of the reflected Homunculus 
spectrum show that the circumstellar extinction along the bipolar axis of the
Homunculus is significantly lower than along our sight line
(Hillier \& Allen \citeyear{HA92_eta}).
In addition infrared observations clearly show that the 
dust is distributed asymmetrically (Smith \etal\ \citeyear{SGH03_IR},
Chesneau \etal\ \citeyear{Ches05_eta_IR}).

We can estimate the optical depth of a resonance line in the wind
using simple scaling laws. Assuming N(H)/N(He)=5 and a solar
mass fraction of Fe, the optical depth can be written in the
form
\begin{eqnarray*}
\tau & =&  1.3 \times 10^5 f  x_l 
               \left( {\lambda \over 0.2\,\mum}\right)
               \left({\Mdot \over 1.0 \times 10^{-3}\,\Msunyr} \right)
               \left( {500\,\kms \over \Vinf} \right)
               \left( {10\,\kms \over V_{th}} \right) \\
                & & \quad\quad \left( { 0.1\arcsec \over r } \right)
               min\left( 1, \sqrt{\pi} (V_{th}/\Vinf)(r/\Rstar) \beta_{eff}^{-1} \right)
\end{eqnarray*}
where $f$ is the oscillator strength, $x_l$ the fraction of the Fe
population in state $l$, and $\beta_{eff}$ is the exponent
which describes the velocity in the wind. For a classic velocity
law $\beta_{eff}=1$, but if we have additional acceleration in
the outer wind $\beta_{eff}$ could be significantly greater than 1.
In the inner regions the optical
depth is determined by the velocity gradient (i.e., 
through the Sobolev approximation), but in the outer region the
static optical depth dominates. It is readily evident that many
of \ion{Fe}{2}\ transitions will have optical depths exceeding unity
at 0.1\arcsec.

Supporting evidence for the importance of resonance scattering comes
from optical observations. First, in the SE lobe we can
resolve the outermost layers of the wind. The spectrum
is dominated by broad permitted and forbidden lines of \ion{Fe}{2}.
Further, near 3000\AA, we see very strong \ion{Fe}{2} emission lines. These most
likely originate by continuum fluorescence.

\begin{figure}
\includegraphics[angle=-90, scale = 0.8]{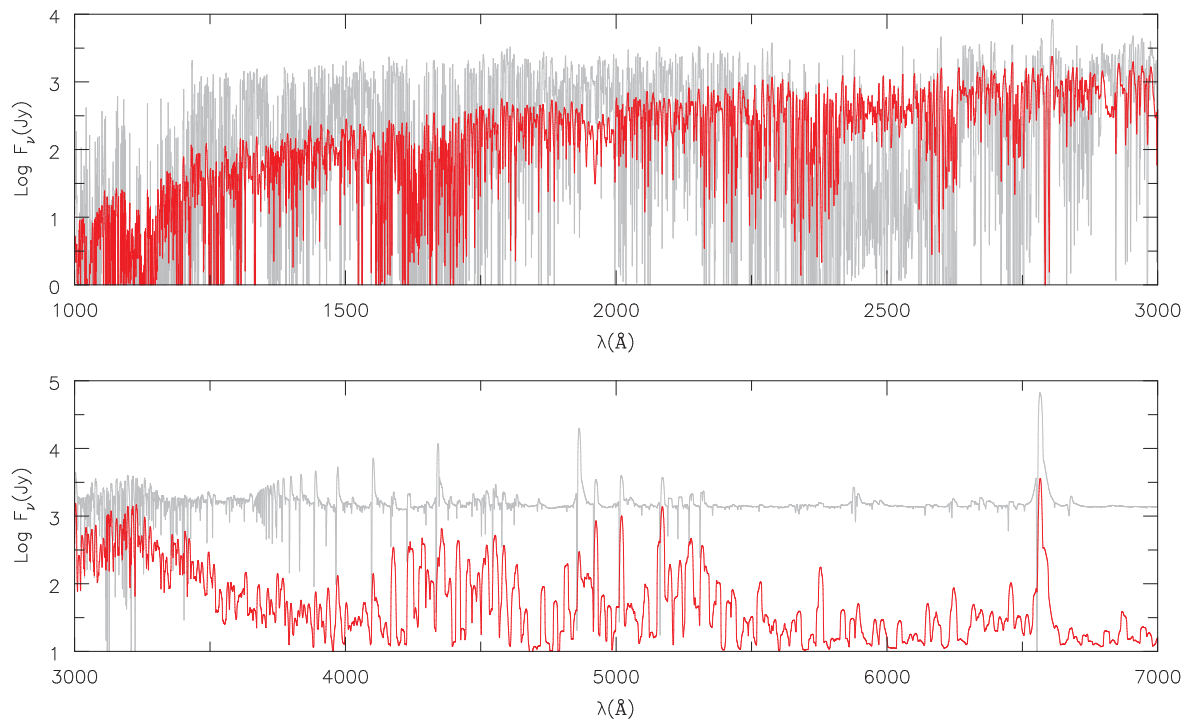}
\caption{Illustration of the spectrum originating inside 0.033\arcsec\
(gray) compared with that originating outside 0.033\arcsec\
(black [red]) as a function of wavelength. In the optical region ($\lambda > 3500$\,\AA)
very little flux originates outside 0.033\arcsec. On the other
hand, in the UV, significant flux originates outside 0.033\arcsec, and this
may dominate what is observed, especially when dust extinction
is taken into account. Independent of the assumptions
about a spatially varying dust extinction, the models indicate that
we are on the verge of resolving \etastar, especially in
the wavelength range from 2500 to 3000\,\AA}
\label{Fig_UV_ext}
\end{figure}

As noted above, we postulate that the UV spectrum originating in the
inner layers (or, more correctly, from small impact parameters) 
is absorbed (and scattered) from our line of sight by dust. 
Thus the UV spectrum we observe is NOT the stellar spectrum ---
rather it is the spectrum that originates outside some impact
parameter.\footnote{By stellar spectrum we mean the spectrum of the
star, and its wind, as would be observed if the star and its winds
were unresolved, and if the Homunculus and other circumstellar ejecta were absent.}
A model comparison with observation needs to specifically address 
this origin at large radii. For simplicity we will assume uniform
extinction centered on the central star, and which completely
blocks the stellar UV light out to some radius (impact parameter). 
In practice the extinction is likely to vary with radial distance from the star, in
azimuth, and depending on the dust properties, with wavelength.  
The spectrum so computed was found to give much better agreement with observation. Indeed
unexplained emission features could now be explained. Note that this does not
explain the anomalous extinction observed in the optical,
since the optical continuum originates at much smaller impact parameters (as
illustrated in Fig.~\ref {Fig_UV_ext})

\begin{figure}
\includegraphics[scale=0.7,angle=-90]{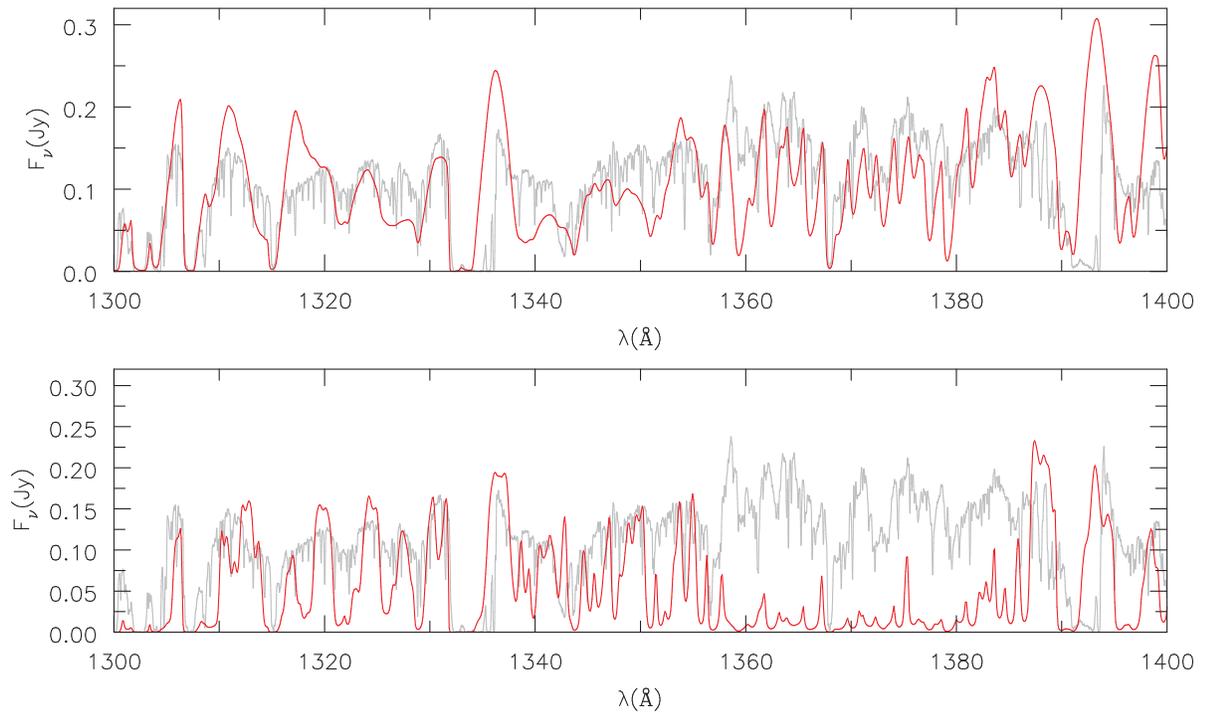}
\caption{Comparison of the spectrum of \etastar\ (gray curve) with the
full model spectrum (bottom panel), and with the spectrum
originating outside 0.033\arcsec (upper panel).  In the latter case there
is much better agreement with observation.}
\label{Fig_UV_1300}
\end{figure}

\begin{figure}
\includegraphics[scale=0.7,angle=-90]{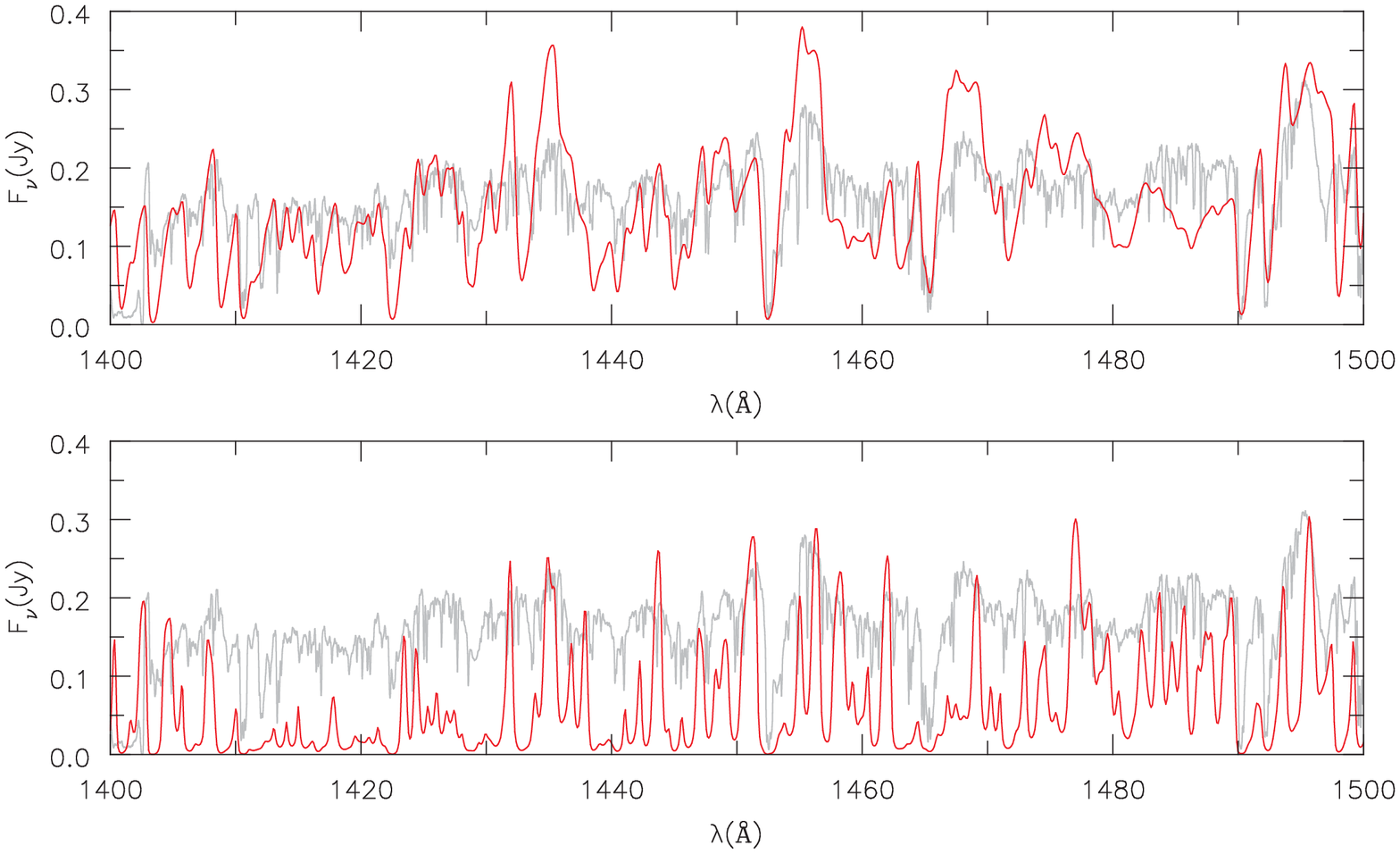}
\caption{As for Fig.~\ref{Fig_UV_1300}. Notice the agreement in
the top panel for the ``flat-topped'' P Cygni profile near
1425\AA. This feature is actually a complicated blend, and has not
been reproduced in any model in which we used the full spectrum
from the central source.}
\label{Fig_UV_1400}
\end{figure}

In Figs. \ref{Fig_UV_1300} \& \ref{Fig_UV_1400} we show a collection of spectra to 
illustrate the agreement/disagreements between model and observation. 
In the bottom panel of each figure we show a direct comparison of the 
observations with the model.  In general there is strong disagreement --- the
theoretical model is too strongly absorbed. However models with lower 
mass loss, and hence higher excitation, still cannot explain the observations.

In the upper panels we compare the spectrum with that arising from outside 
0.033\arcsec.  There is much better qualitative agreement between the observed 
and predicted spectra.  
In particular, notice how the P~Cygni emission feature near 1425\AA\ is
reproduced in the models. The emission feature is actually a complicated 
blend arising from many lines. No integrated spectrum from
a single mass-loss rate model could reproduce this feature.
Interestingly, as noted earlier, the spectrum originating outside 
0.033\arcsec\ also provides a better fit to the 
FUSE spectrum (Fig.~\ref{Fig_fuse_prim}).

\begin{figure}
\includegraphics[scale = 0.7]{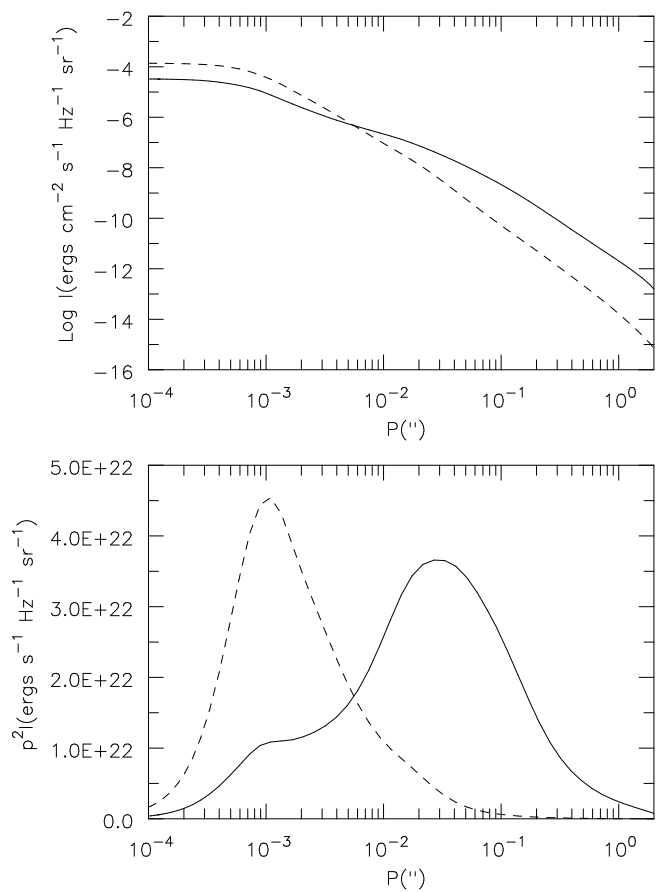}
\caption{Illustration of how the size of the star varies with wavelength.
In the top panel we have plotted the specific intensity as a function of 
impact parameter ($p$),
averaged over 2 wavebands (2700-3000\AA [solid line]; and 
5500-6500\AA [dashed line]). In both
bands the specific intensity peaks on small scales, and there is a well defined
stellar ``core''. In the lower panel we have scaled the
specific intensity by $p^2$. Thus the area under the curve [$\log p$ versus $p^2 I(p)$]
is proportional to the flux. In the UV the star is extremely extended due
to scattering by \ion{Fe}{2} bound-bound transitions.}
\label{Fig_IP}
\end{figure}

The choice of 0.033\arcsec\ was (somewhat) arbitrary, but for ``small'' 
changes (see Fig.~\ref{Fig_IP}) the conclusions and
spectral comparisons are still valid. While 0.033\arcsec\ may seem unduly small,
some support for such a value can be gleaned from the Weigelt blobs. 
Since the Weigelt blobs suffer much less extinction than the central
star, the coronagraph, at least on the NW side extends $< 0.2$\arcsec.

\begin{figure}
\includegraphics[scale = 0.7]{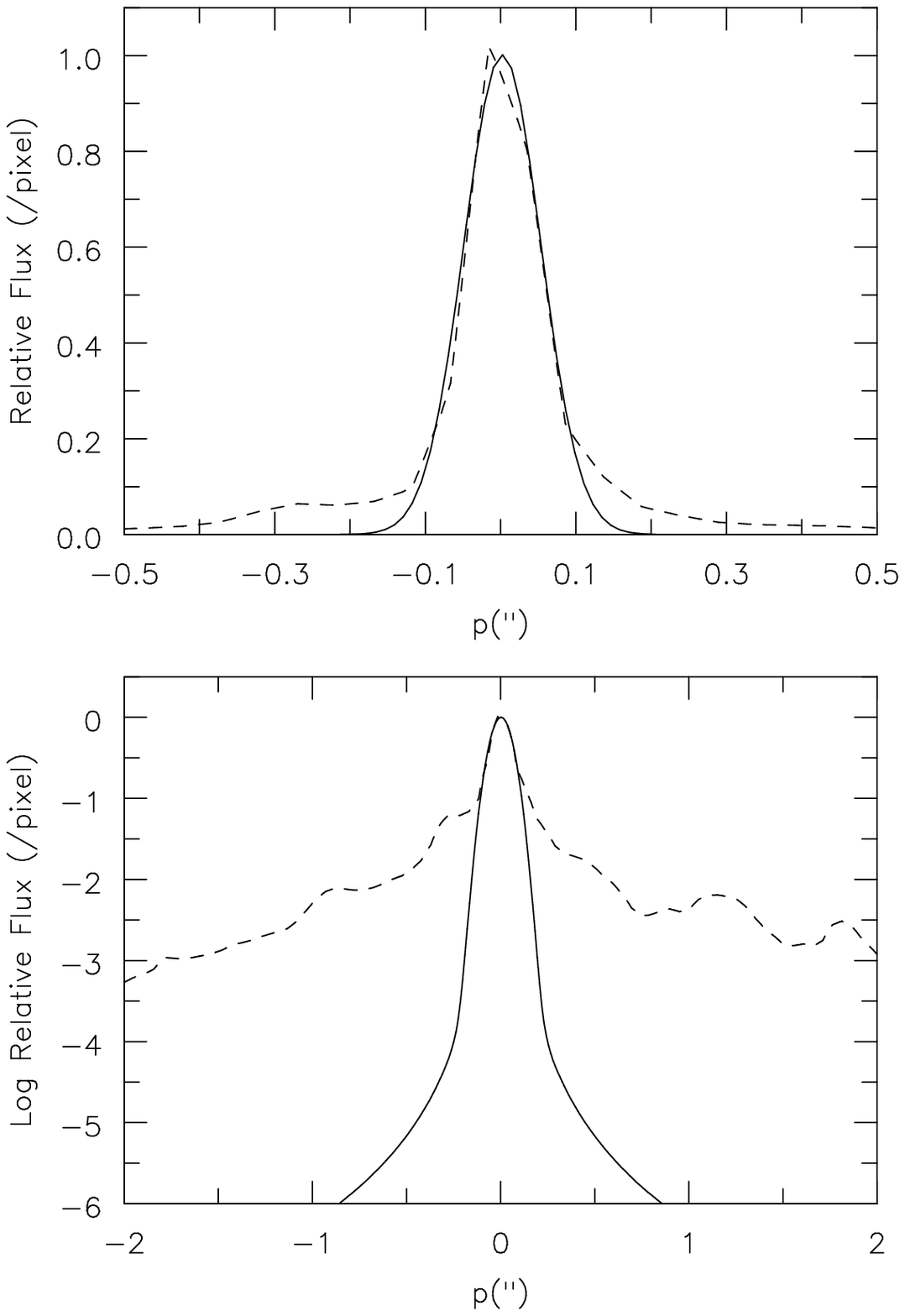}
\caption{Illustration of the flux variation across a 0.1\arcsec\ aperture
using the data of 04-Jul-2002 (wavelength range 2900-3000\AA; dashed line). 
Shown for comparison is the expected theoretical
distribution computed assuming a FWHM of 0.12\arcsec\ for the telescope/instrumental
point-spread function (scaled so that the heights match). 
Despite the extension of the star, the core of the profile is dominated by the 
telescope/instrument point-spread function. In the outer regions the
observed flux distribution lies well above the model predictions ---
this extended halo is presumably due to dust scattering.
For simplicity we used a 1D convolution.}
\label{Fig_core}
\end{figure}

In Fig.~\ref{Fig_UV_ext} we provide an illustration of the extension
of the star as a function of wavelength. The extension is strongly
wavelength dependent, and is largest in the UV, particularly from 2500 to 3000\,\AA.
This figure indicates that at UV wavelengths, particularly from 
2500 to 3000\AA, a significant fraction of the UV flux originates outside
0.033\arcsec. While the source is extended we should note that
the model predicts that the star should, in the absence of
dust scattering, still have a well defined ``stellar core''.
This is illustrated in Figure~\ref{Fig_IP}. In the upper panel we see that the
star has a well defined core, although the size of the core varies
with wavelength. In the lower panel, we take into account the
integration over area such that the area under the curve, as illustrated,
is proportional to the observed flux. In the UV (2700-3000\AA) the star
is roughly 30 times larger than it is in the optical (5500-6500\AA). 
The smaller peak in  the UV probably reflects regions between lines.
Despite the extended UV structure, the observed point spread function
along the slit of the CCD detector will be dominated by the
point spread function of the telescope and instrument 
(Fig.~\ref{Fig_core}). The model does not explain the very extended
UV emission, which presumably arises from dust scattering in the
Homunculus.

Detailed quantitative illustrations of the extended UV halo
are provided by Smith \etal\ \cite{SMG04_UV1}.  In the F220W and F250W filters
there is an extended ``bright'' halo around the point source, not seen in
the optical filters, which extends out to about 0.5\arcsec. The extended
UV emission can be characterized by the amount of flux emitted in a 0.1\arcsec\
aperture centered on the star relative to that emitted in 
a 3.2\arcsec\ diameter aperture. For the F550M,  F330W, F250W, and
F220W filters the fractions are 27\%, 14\%, 9.5\%, and 8.6\%, respectively.

\subsection{The UV spectrum before and during the event}

A detailed paper on the variability of the UV spectrum before, during,
and after the 2003 minimum will be provided elsewhere. Here we provide only
a brief summary. The data sets covering March 2000 to
July 2004 are very similar in the UV.
In the wavelength region covered by E140M (1150 to 1700\AA)
there is an overall increase in flux with time --- 
the data set obtained in July 2002 is approximately 40\% brighter 
than that in March 2000. Qualitatively, however, the spectra
are similar, and the changes are too small to affect any conclusions
drawn in this paper. The blue edge of the P~Cygni profiles
is remarkably constant, while significant changes are seen on
the red side of the absorption profile (weaker in later data sets),
and there are some changes in the emission line strengths.

During the UV event,
the \ion{Fe}{2}\ forest in the UV becomes much more prominent. In some
regions around 2200\AA, it is difficult to trace the star in the
MAMA aperture. In addition the UV emission is extended,
it is not symmetrical about the star, and we no longer have
a well defined point source. The enhancement of the \ion{Fe}{2}\
forest was expected, since it was known that the P~Cygni absorption
components on the \ion{Fe}{2}\ optical lines were much stronger
near and during the minimum. Two simple scenarios can be used
to explain the absorption profile variability. In the first, a
shell is ejected which reduces the UV flux to the outer wind
leading to a lower iron ionization and hence enhanced P Cygni
profiles. In the second scenario, UV flux from the companion star
is blocked from reaching the outer wind material 
when the star is near periastron.

\section{The Optical Outer Wind Spectrum}
\label{Sec_out_wind_spec}

Using STIS on the HST it is possible to resolve the outer wind.
Potentially this could provide a wealth of information on the nature
of the stellar wind. Unfortunately the analysis of the spectra 
will be difficult:
\begin{enumerate}
\item
   The extinction, and the shape of the extinction law, is unknown.
   It varies with location across the wind on the sky, and
   within the wind.
\item
  The outer wind spectrum is contaminated by the stellar (inner wind)
 spectrum
  due to dust scattering (Hillier \& Allen \citeyear{HA92_eta}; 
   Smith \etal\ \citeyear{SDG03_eta_lat}).
  While the inner wind spectrum 
  can be readily identified, it is not easily removed since the
  EWs of the scattered wind lines in the spectrum are often lower
  than those measured on the central source. The reason for this is unknown,
  although it could simply be a consequence of the wind asymmetry.
  Alternatively it could be related to the occurrence of dust within the stellar wind,
  or the properties of the intrinsic coronagraph.
\item
   Instrumental artifacts (e.g., ghosts)
   can cause spurious features which can mimic the
   dust scattered spectrum (e.g, Hill \citeyear{Hil00_ghost}, Martin
   \citeyear{Mar04_ghost}).  The location of these features in 
   long-slit Homunculus spectra will depend on the HST slit orientation.
   The ghost in CCD spectra are due to window in front of the CCD housing.
   This is not a problem with the MAMAs although there is a much smaller ghost in
   the FUV MAMA, closer to the point source, due to the photocathode being on the 
   window a short distance from the microchannel plate.

\end{enumerate}

\begin{figure}
\includegraphics[scale=0.7,angle=-90]{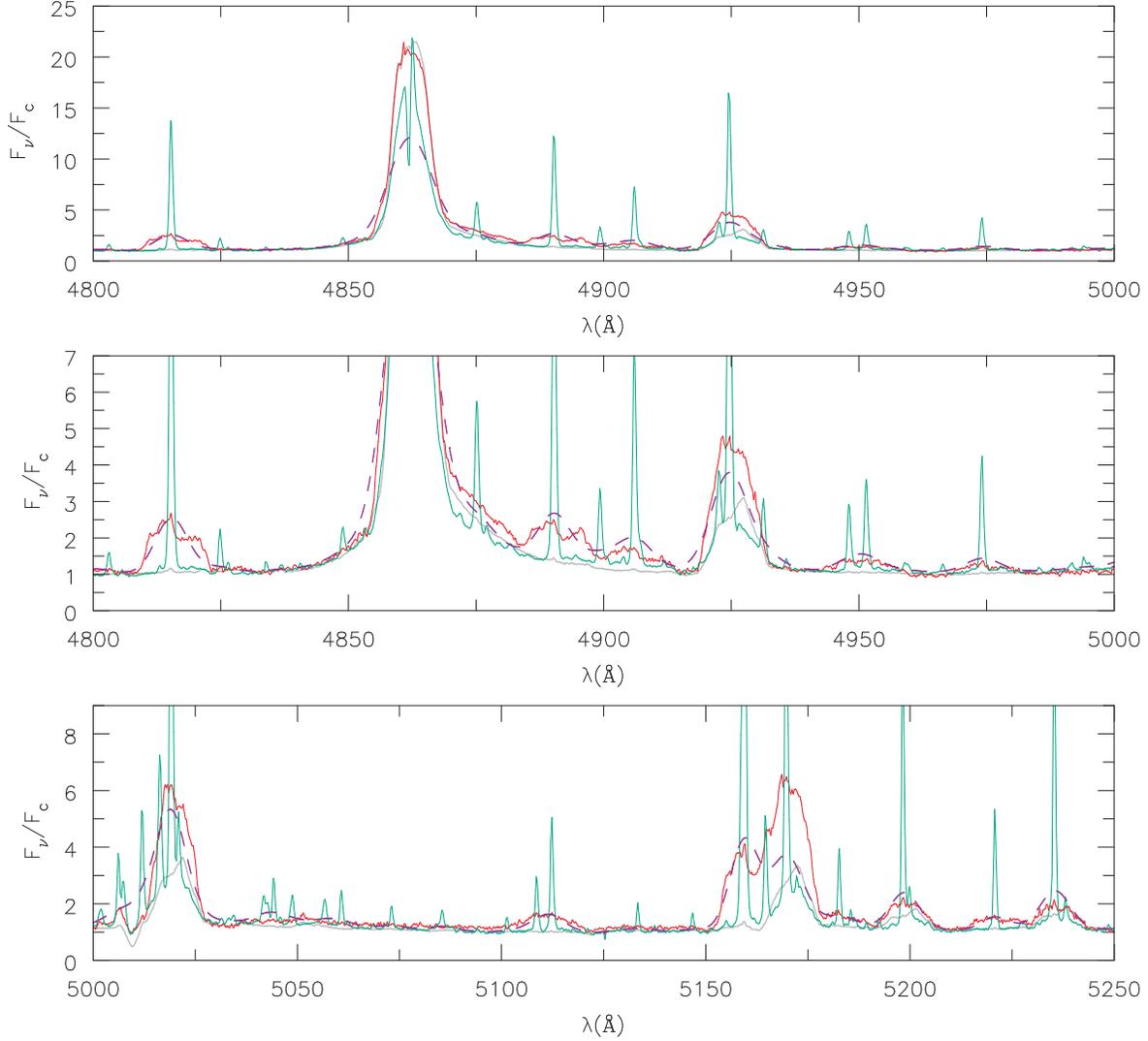}
\caption{Illustration of the flux as observed 0.2\arcsec\ from the
central source. Four curves are shown: solid red -- 
wind spectrum 0.2\arcsec\ NE of \etastar\ but in the SE lobe; 
gray --- stellar spectrum; green spectrum with narrow lines --- spectrum of
the Weigelt blob; purple dashed --- spectrum of Weigelt blob smoothed so that
the FWHM of the lines are similar to that of the wind spectrum.
The continuum have been normalized to unity.
Notice that at the 0.2\arcsec\ location H$\beta$ has a similar strength to that on
the star, but that in the Weigelt blob spectra it is significantly
weaker. Despite the predominance of dust scattering, H$\beta$ is generally
seen to be weaker in spectra taken off the central star (e.g., in the homunculus).
A pure [\ion{Fe}{2}] line ($\lambda 4815$; 20F), which is formed at
0.2\arcsec, can be readily identified.}
\label{Fig_rw_a}
\end{figure}

\begin{figure}
\includegraphics[scale=0.7,angle=-90]{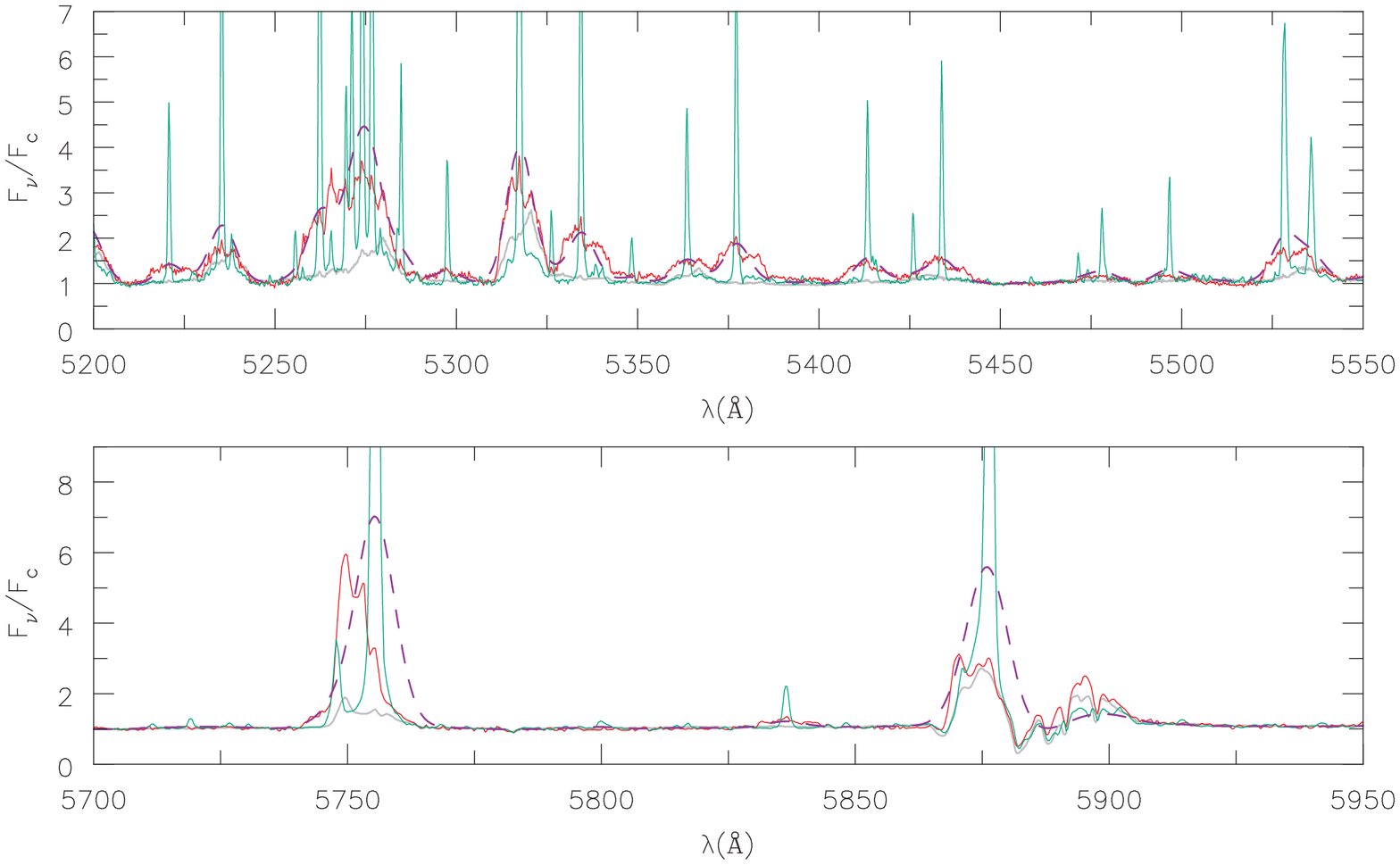}
\caption{As for Fig.~\ref{Fig_rw_a}. Particularly striking is the asymmetric
[\ion{N}{2}] $\lambda$5754 line profile in the wind spectrum.}
\label{Fig_rw_b}
\end{figure}

On the NW side the wind spectrum is not easily discerned.
A detailed analysis of the outer wind spectrum will be left to
a future paper. In Figs.~\ref{Fig_rw_a} and \ref{Fig_rw_b}
we show a comparison of the wind spectrum with that
of the Weigelt blobs. The wind spectrum was obtained
at 0.2\arcsec\ NE of the star but still in the SE lobe
(PA 69$^\circ$) on July 4, 2002. 
To facilitate the comparison, we have 
broadened the Weigelt blob spectrum so that the line widths
correspond roughly to that seen in the stellar wind. 
Several similarities can be seen with the Weigelt blob spectra. However,
there are also important differences.

\begin{enumerate}

\item
The stellar spectrum does not show the high excitation lines (e.g.,
[\ion{Ne}{3}] $\lambda 3869$; [\ion{Fe}{3}] $\lambda \lambda$4658, 
4701; [\ion{Ar}{3}] $\lambda 7136$) seen in the spectra of the
Weigelt blobs (although not at minimum). However blue shifted components
of these lines can be seen in the wind spectrum.

\item
Many of the forbidden \ion{Fe}{2}\ lines are also seen in the
wind spectrum, although they are broadened. The severe blending
is seen by direct comparison with the Weigelt blob spectrum.

\item
At longer wavelengths (i.e., far red), the [\ion{Fe}{2}] spectrum is not as
prominent. This can be attributed to several possible causes: 
the general weakening of the [\ion{Fe}{2}] lines, the strong scattered continuum 
which makes the lines more difficult to discern, and
the absence of lines pumped by Ly$\alpha$.

\item
In the 3000 to 3300\AA\ region many strong
broad lines can be seen. Some of these have counterparts in the
star, but NOT in the Weigelt blobs. We attribute these lines
to continuum fluorescence.

\item
Because of line blending, line profiles are difficult to ascertain. 
Many lines are fairly symmetric (e.g, \ion{Fe}{2} $\lambda 4923$),
and exhibit a parabolic profile with a slight flattening at the top.
On the other hand, some lines (e.g., [\ion{N}{2}] $\lambda 5754$) show
very asymmetric profiles --- the [\ion{N}{2}] $\lambda 5754$ profile appears
blue shifted, and is quite asymmetric. This profile may 
arise from a distinct emitting region --- the profile appears
to be related to the blue shifted components of [\ion{Fe}{3}]
seen at this location (see item 1), and which have been identified in
ground-based spectra (Zanella \etal\ \citeyear{ZWS84_shell}).

\end{enumerate}

Particularly noteworthy is that broad forbidden lines indicate a
terminal velocity of 500\,\kms\, similar to that derived from the stellar
spectrum. The outer wind features give rise to the broad base of
the Prussian helmets seen in ground-based spectra
(Hillier \& Allen \citeyear{HA92_eta}).
The resolution of the stellar wind at optical wavelengths,
and the detection of broad [\ion{Fe}{2}] lines, lends support to 
our interpretation of the UV, and to our model for $\eta$ Carinae. 

\section{Conclusion}
\label{Sec_conc}

A fundamental prerequisite to understanding the reason for the
ejection of the Homunculus is to determine the fundamental properties
of the primary star. Unfortunately the primary star is shrouded in a
dense wind. In addition, dust extinction does not allow an uninterrupted
view of the primary star, and its wind.

The UV spectrum of Eta Carinae is dominated by low ionization lines, 
many exhibiting P Cygni profiles. Some of the strongest lines that 
can be readily identified include \ion{C}{2} $\lambda$ 1335 (UV1),
\ion{Si}{2}  $\lambda\lambda$1304, 1309 (UV3); $\lambda$ 1264 (UV4), 
$\lambda\lambda$ 1527, 1533 (UV 2); $\lambda\lambda$  1808, 1817 (UV1),
\ion{S}{2} $\lambda\lambda$ 1250, 1253 (UV1), 
\ion{Al}{2} $\lambda$1671, 
\ion{N}{1}\ $\lambda\lambda$ 1493, 1495 (UV4), 
\ion{Mg}{2} $\lambda\lambda$ 2796, 2803, as well as numerous \ion{Fe}{2} lines.
Higher excitation lines due to \ion{Al}{3} $\lambda\lambda$1855, 1863,
and probably \ion{Si}{4} $\lambda\lambda$1394, 1403 can also
be identified. The identification of \ion{C}{4}\ $\lambda\lambda$ 
1548, 1552 is uncertain because of severe line blending.

We have shown that we do not directly observe the star and
its wind in the UV --- rather, because the inner regions are occulted by dust, 
we only observe the spectrum created in the outer regions (i.e., at
large impact parameters) of the stellar wind. 
This helps partially explain the flatness of the
UV extinction law --- the flatness does not simply reflect the
properties of the dust. Importantly, it indicates that we cannot simply
use the observed UV spectrum to determine the properties of the dust
causing the circumstellar extinction. Further the dust in the densest knots
will suffer greater shielding from UV radiation, and will have different 
properties compared to the dust located in other regions. This affects our
ability to understand the excitation of the Weigelt blobs.
The results reinforce the belief that our view of Eta Carinae is
biased. The view from other directions would be significantly
different. It is still probable the ionizing flux from the
companion star is also influencing the observed spectrum.
The high hydrogen column density towards the central star 
($\log \, $N(H)$=22.5$) 
is confirmed from analysis of the Ly$\alpha$ profile in HST MAMA data.

We show that the FUSE spectral region is, in principle, the best
wavelength region to detect directly the primary star. However,
no direct evidence of the companion star, with the properties
indicated by X-ray studies (i.e, $\Mdot \sim 10^{-5}\,\Msunyr$, and $\Vinf=3000\,$\kms)
is seen in current FUSE data, or in MAMA data. This may be partially a consequence
of reprocessing of the companion light by the dense wind of the
primary. Alternatively, it may indicate the parameters of the O star,
as inferred from the X-ray and Weigelt blob analyses, are incorrect.

The best fit to the FUSE data is obtained using the spectrum
for the model of the primary originating beyond 0.033\arcsec.
The fit, not surprisingly, is far from perfect. As our
study has shown modeling of the UV spectra of Eta Car is extremely
difficult. It is necessary to allow for the extended nature
of the UV emitting region and occulation by dust. Further, the companion
will modify the ionization structure of the wind, and while
the companion is probably the dominant light source, in the FUSE spectral region,
its spectrum will be modified by the dense wind of the primary. 
Finally, there is extensive evidence for an axi-symmetric wind  which 
also needs to be allowed for in future modeling.


The terminal velocity of the primary's wind lies between 500 and 600\,\kms\
with values near the lower end preferred. This range is
consistent with that determined by HDIG from analysis of the
optical spectrum. Surprisingly, we find no convincing evidence for higher
velocity components. 

With the STIS on the HST we have resolved the stellar
wind of Eta Carinae. Broad \ion{Fe}{2}\ emission lines
are observed directly. These broad wind lines can be seen at a distance
of 0.2\arcsec\ (and beyond) from the central source, and also
indicate a wind terminal velocity of approximately 500\,\kms.
The wind spectrum shows some similarities to the spectra of
B \& D Weigelt blobs, but also shows some marked differences
in that high excitation lines, and lines pumped
by Ly$\alpha$, are not seen.

\blankline
The observations were made with the NASA/ESA Hubble Space
Telescope under HST-GO and STIS-GTO programs through the STScI
under NAS5-26555 and with the NASA/CNES/CSA Far Ultraviolet Spectroscopic Explorer, which is 
operated for NASA by The Johns Hopkins University under NASA contract 
NAS5-32985.  T.R.G., G.S., R.C.I. \& D.J.H. acknowledge support from the 
FUSE Guest Investigator program. The HST Treasury project is supported
by NASA programs GO-9420 and GO-9973.
K. W. acknowledges support by the state of North Rhine-Westphalia
(Lise Meitner fellowship). We would also like to thank K. Davidson
and R. M. Humphreys for providing useful comments on an earlier
draft of the manuscript.

\def\SA_BCM#1{in  NATO ASI Seri. C 341, 
Stellar Atmospheres: Beyond Classical Models,
ed. L. Crivellari, I. Hubeny \& D. G. Hummer (Dordrecht: Kluwer), #1}

\def\ASP#1{in ASP Conf. Ser. 22, Nonisotropic and Variable 
Outflows from Stars,
ed. L. Drissen, C. Leitherer, \& A. Nota (San Francisco: ASP) p.~#1}

\def\prco#1{in Polarized Radiation of Circumstellar Origin,
ed. G. V. Coyne, A. M. Magalh\~aes, A. F. J. Moffat, 
R. E. Schulte-Ladbeck, \& S. Tapia (Vatican Observatory),~#1}

\def\plbv#1{in IAU Colloq. 113, Physics of Luminous Blue Variables,
ed. K. Davidson, A. F. J. Moffat, \& H. J. G. L. M. Lamers 
(Dordrecht: Kluwer),~#1}
       
\def\lbv#1{1997, in ASP Conf. Ser. 120, 
Luminous Blue Variables: Massive Stars in Transition,
ed. A. Nota, \& H.J.G.L.M. Lamers (San Francisco: ASP),~#1}

\def\IAU70#1{1976, in IAU Symp. 70, Be and Shell Stars, 
ed. A. Slettebak (Dordrecht: Reidel)~#1}

\def\etaworkshop#1{in ASP Conf. Ser. 179, 
Eta Carinae at the Millenium, ed. J.A. Morse, R.M. Humphreys, \&
A. Damineli (San Francisco: ASP),~#1}

\def\Hveneta#1{in ASP Conf. Ser. 242, Eta Carinae and Other Mysterious
Stars: The Hidden Opportunities of Emission Line Spectroscopy,
ed. T.R. Gull, S. Johansson, \& K. Davidson (San Francisco: ASP),~#1}

\end{document}